\documentclass[onecolumn,preprint]{revtex4-1}

\usepackage{amsmath,amsfonts,amssymb}
\usepackage{graphicx}
\usepackage{dcolumn}
\usepackage{bm}
\usepackage{xcolor}
\usepackage{float}

\definecolor{arsenic}{rgb}{0.23, 0.27, 0.29}
\definecolor{blue-violet}{rgb}{0.54, 0.17, 0.89}
\definecolor{britishracinggreen}{rgb}{0.0, 0.26, 0.15}
\definecolor{ferngreen}{rgb}{0.31, 0.47, 0.26}
\definecolor{forestgreen}{rgb}{0.13, 0.55, 0.13}
\definecolor{limegreen}{rgb}{0.2, 0.8, 0.2}

\newcommand{\ket}[1]{\vert\mskip3mu#1\mskip3mu\rangle}
\newcommand{\braket}[2]{\langle\mskip3mu#1\mskip3mu\ket{#2}}
\newcommand{\braopket}[3]{\langle\mskip3mu#1\mskip3mu\vert\mskip3mu#2\mskip3mu\vert\mskip3mu#3\mskip3mu\rangle}
\newcommand{\threejm}[6]{ \left(\begin{array}{ccc} #1 & #3 & #5\\
                                                   #2 & #4 & #6 \end{array} \right)}

\newcommand{\bmR}{{\bm R}}

\newcommand{\bmJ}{{\bm J}}
\newcommand{\bmj}{{\bm j}}

\newcommand{\rme}{\mathrm{e}}
\newcommand{\rmo}{\mathrm{o}}

\newcommand{\bbR}{{\mathbb{R}}}

\begin{document}

\title{State-to-state rovibrational transition rates for CO$_2$ in the bend mode in collisions with He atoms}

\author{Taha Selim}
\email{tselim@science.ru.nl}
\affiliation{Theoretical Chemistry \\
Institute for Molecules and Materials, Radboud University \\
Heyendaalseweg 135, 6525 AJ Nijmegen, The Netherlands}

\author{Ad van der Avoird}
\affiliation{Theoretical Chemistry \\
Institute for Molecules and Materials, Radboud University \\
Heyendaalseweg 135, 6525 AJ Nijmegen, The Netherlands}

\author{Gerrit C. Groenenboom}
\email{gerritg@theochem.ru.nl}
\affiliation{Theoretical Chemistry \\
Institute for Molecules and Materials, Radboud University \\
Heyendaalseweg 135, 6525 AJ Nijmegen, The Netherlands}

\date{\today}

\begin{abstract}
Modeling environments that are not in local thermal equilibrium, such as
protoplanetary disks or planetary atmospheres, with molecular
spectroscopic data from space telescopes requires knowledge of the rate
coefficients of rovibrationally inelastic molecular collisions. Here, we
present such rate coefficients in a temperature range from 10 to 500\,K
for collisions of CO$_2$ with He atoms in which CO$_2$ is (de)excited in
the bend mode. They are obtained from numerically exact coupled-channel
(CC) calculations as well as from calculations with the less demanding
coupled-states approximation (CSA) and the vibrational close-coupling
rotational infinite-order sudden (VCC-IOS) method. All of the
calculations are based on a newly calculated accurate \textit{ab initio}
four-dimensional CO$_2$-He potential surface including the CO$_2$ bend
($\nu_2$) mode. We find that the rovibrationally inelastic collision
cross sections and rate coefficients from the CSA and VCC-IOS
calculations agree to within 50\% with the CC results at the rotational
state-to-state level, except for the smaller ones and in the low energy
resonance region, and to within 20\% for the overall vibrational
quenching rates except for temperatures below 50\,K where resonances
provide a substantial contribution. Our CC quenching rates agree with
the most recent experimental data within the error bars. We also
compared our results with data from Clary \textit{et al.} calculated in
the 1980's with the CSA and VCC-IOS methods and a simple atom-atom model
potential based on \textit{ab initio} Hartree-Fock calculations and
found that their cross sections agree fairly well with ours for
collision energies above 500 cm$^{-1}$, but that the inclusion of long
range attractive dispersion interactions is crucial to obtain reliable
cross sections at lower energies and rate coefficients at lower
temperatures.
\end{abstract}

\maketitle

\section{Introduction}
\label{sec:intro}
The evolution of interstellar molecular clouds to protostars to
protoplanetary disks to planetary systems can be followed by observing
spectra emitted or absorbed by small molecules, such as CO, CO$_2$, HCN,
C$_2$H$_2$, etc. \cite{roueff:13}. Microwave spectra are generated by
rotational transitions in these molecules, infrared spectra by
rovibrational transitions. The shapes of the lines in the spectra depend
on the rotational and vibrational level populations, and these
populations are determined both by radiative transitions and by
transitions caused by molecular collisions with abundant species:
hydrogen and helium atoms, H$_2$ molecules, and electrons. The
information from the spectra is used by astronomers in modeling the
processes taking place in the various stages of the evolution. Two
situations are distinguished: (1) environments from which the spectra
originate are in local thermal equilibrium (LTE) and the line shapes in
the spectra only depend on the temperature, and (2) the spectra
originate from non-LTE environments. In the first situation it is
sufficient to know the Einstein $A$ and $B$ coefficients for spontaneous
and stimulated emission and absorption. In the non-LTE case one also
needs to know state-to-state transition rate coefficients from molecular
collisions. This paper is concerned with the latter case, and in
particular with the calculation of collisional transition rates from
first principles. Cross sections and rate coefficients for rotationally
inelastic collisions have already been calculated for several
astronomically relevant molecular systems including CO$_2$-He
\cite{roueff:13,palluet:22}, but the additional inclusion of vibrational
(de)excitations is more demanding. In earlier work
\cite{song:13a,song:15a,walker:15,song:15b} we studied rovibrational
transitions in CO in collisions with H atoms and rovibrational
transitions in the stretch modes of CO$_2$ in collisions with He atoms
\cite{selim:21,selim:22}. Here,  we investigate the bend mode of CO$_2$
in collisions with He atoms.

Spectra of the CO$_2$ bend mode in the 15 $\mu$m range originating from
different astronomical environments were observed by the Infrared Space
Observatory (ISO) \cite{boonman:03,boonman:03a} and the Spitzer Space
Telescope \cite{sonnentrucker:08} and, recently, also by the James Webb
Space Telescope (JWST) \cite{grant:23}. Another process in which
collisions of CO$_2$ in the stretch and bend modes with He atoms are
important is the CO$_2$ laser action
\cite{patel:64,offenberger:70,clary:83,kasner:08}. Furthermore, the
CO$_2$ bend mode is relevant because it yields the dominant contibution
to the terrestial greenhouse effect \cite{allen:79,lopez:92}, and
collisions with (oxygen) atoms in the mesosphere are important as well
\cite{feofilov:12}. Finally, CO$_2$ occurs in the atmospheres of other
planets and exoplanets \cite{jwst:23} and its spectra are important in
modeling these atmospheres.

The CO$_2$ molecule has three vibrational modes: a twofold degenerate
bend mode with experimental frequency 667~cm$^{-1}$, a symmetric stretch
mode at 1333~cm$^{-1}$, and an asymmetric stretch mode at 2349~cm$^{-1}$
\cite{shimanouchi:72}. Here, we concentrate on the bend mode. In their
pioneering theoretical work on rate coefficients for vibrational
transitions in CO$_2$ induced by collisions with rare gas (Rg) atoms
Clary \textit{et al.}
\cite{clary:81,clary:81b,clary:82,clary:83,banks:87b,wickham:87b} also
investigated the bend mode. In scattering calculations they used the
coupled-states approximation (CSA), as well as the VCC-IOS method, a
vibrational coupled-channel (CC) method for the vibrations, combined
with the infinite-order sudden (IOS) approximation for the rotations.
Although they included the rotational states both in their VCC-IOS
calculations \cite{clary:83} and in their CSA calculations
\cite{banks:87b}, they only provided some illustrative data for
state-to-state rovibrational transitions. And they used a model
potential based on \textit{ab initio} self-consistent field (SCF)
calculations, which can nowadays be calculated much more accurately.
Experimental data are available only for overall vibrational transition
rates \cite{lepoutre:79,siddles:94}. The more advanced models currently
being developed by astronomers \cite{bosman:17,bosman:19} and the
availability of data from JWST require rovibrational state-to-state
collisional rate coefficients, which we here present.

Section~\ref{sec:pot} describes the \textit{ab initio} calculation of
the 4-dimensional CO$_2$-He intermolecular potential depending on the
CO$_2$ bend coordinate. Also the analytical representation of the
potential is defined and the potential is illustrated. The bend mode of
a linear molecule like CO$_2$ is twofold degenerate and generates
vibrational angular momentum, which makes the theory more complicated
than it is for the stretch modes or the bend modes in nonlinear
molecules. It is outlined in Sec.~\ref{sec:theory} as part of the
CO$_2$-He scattering approach. In Sec.~\ref{sec:results} we present and
discuss our results and compare the overall vibrational transition rates
with the available experimental data and with the results of Clary
\textit{et al.} Section~\ref{sec:concl} summarizes our conclusions.

\section{Coordinates and frames}
The Jacobi scattering vector points from the center-of-mass of CO$_2$ to
the helium atom. Its Cartesian coordinates with respect to a space-fixed
(SF) frame are given by the column vector $\bmR$. The spherical polar
coordinates of this vector are $(R,\Theta,\Phi)$, with $R=|\bmR|$. In
the scattering calculation we express the wave function in a two-angle
embedded body-fixed (BF) frame with the vector $\bmR$ as its $z$-axis,
which isdefined by the rotation matrix $\bbR(\Phi,\Theta,0)$. This
matrix is written in $zyz$-Euler angle parameterization using the active
rotation convention, see Biedenharn and Louck \cite{biedenharn:81}
page 23, i.e.,
\begin{equation}
  \label{eq:BF}
  \bmR = \bbR(\Phi,\Theta,0)\bmR^\mathrm{BF} = \bbR_z(\Phi)\bbR_y(\Theta)
    \bmR^\mathrm{BF},
\end{equation}
where $\bmR^\mathrm{BF}=(0,0,R)^T$ are the BF coordinates of the vector $\bmR$.
The rotation matrices $\bbR_z(\Phi)$ and $\bbR_y(\Theta)$ represent
rotations around the $z$- and $y$-axes, respectively, see, e.g.,
Ref.~\cite{biedenharn:81} or Eq.~(5) in \cite{dhont:04}.

We also define a molecule-fixed (MF) frame, which has its $z$-axis
parallel to the vector that connects the two O atoms and which has the bent
CO$_2$ molecule in the $xz$-plane.
The MF coordinates of the Jacobi vector are related to its BF
coordinates through
\begin{equation}
  \label{eq:MF}
  \bmR^\mathrm{BF} = \bbR(\alpha,\beta,\gamma)\bmR^\mathrm{MF},
\end{equation}
where $\bbR(\alpha,\beta,\gamma)$ defines the MF frame with respect to the
BF frame $\bbR^\mathrm{BF}(\Phi,\Theta,0)$. Inverting this equation gives
\begin{equation}
   \bmR^\mathrm{MF} = \bbR(\alpha,\beta,\gamma)^T \bmR^\mathrm{BF}
   = R\begin{pmatrix}
       -\sin\beta\cos\gamma\\ \sin\beta \sin\gamma \\ \cos\beta
   \end{pmatrix}.
\end{equation}
The superscript $T$ on the rotation matrix means transpose, which
gives its inverse, since the matrix is orthonormal. The angle
$\alpha$ drops out of the equation because the vector $\bmR^\mathrm{BF}$
is invariant under rotations around the BF $z$-axis.
This equation shows that the angles $\beta$ and $\gamma$ are related
to the spherical polar angles $(\beta',\gamma')$ of the vector $\bmR^\mathrm{MF}$
by $\beta'=\beta$ and $\gamma'=\pi-\gamma$. In Sec.~\ref{sec:pot} we use
these angles to define the potential.

Combining Eqs.~(\ref{eq:BF}) and (\ref{eq:MF})
we find that the MF frame is given with respect to the SF frame by
\begin{equation}
  \bbR(\Phi,\Theta,0)\bbR(\alpha,\beta,\gamma) =
  \bbR(\Phi,\Theta,\alpha)\bbR(0,\beta,\gamma),
\end{equation}
where on right-hand-side we have a three-angle embedded BF
frame $\bbR(\Phi,\Theta,\alpha)$ and a two-angle embedded MF frame
$\bbR(0,\beta,\gamma)$. We use these frames in the VCC-IOS calculations,
see Sec.~\ref{sec:VCCIOS}.

\section{Four-dimensional CO$_2$(bend)-He potential}
\label{sec:pot}
The coordinates in the CO$_2$-He potential
$V(\tilde{Q},R,\beta',\gamma')$ are the spherical polar coordinates
$(R,\beta',\gamma')$ of the helium atom in the MF frame defined above
and the dimensionless amplitude
$\tilde{Q}$ of the bend mode in the harmonic approximation, which is defined
in Sec.~\ref{sec:monomer} below; the classical turning points are
$\tilde{Q}=\pm 1$ for $v=0$ and $\tilde{Q}=\pm \sqrt{3}$ for $v=1$.

The potential was calculated with the \textit{ab initio} coupled-cluster
method with single and double excitations and perturbative triples,
CCSD(T), using the MOLPRO package \cite{molpro}. The basis was the
augmented triple-zeta correlation-consistent polarized (aug-cc-pVTZ)
basis of Dunning and coworkers \cite{dunning:89}, supplemented with a
set of $3s3p2d1f$ midbond functions. These midbond functions were placed
on the intersection of the vector $\bm{R}$ and an ellipsoid around
CO$_2$, as described in Refs.~\cite{selim:21,groenenboom:03}. This
ellipsoid is chosen such that it corresponds to the midpoint of $\bm{R}$
at T-shaped configurations with $\beta' = 90^\circ$ and to the midpoint
of the vector connecting He with the nearest O atom in linear
configurations with $\beta' = 0^\circ$ and $180^\circ$. This choice of
the location of midbond functions and the use of geometry dependent
exponents prevents overcompleteness of the basis, especially in the
short range. In all CCSD(T) calculations the $T_{1}$ diagnostic was less
than $0.018$, which indicates the reliability of the CCSD(T) method. The
interaction energies were computed using the Boys and Bernardi
\cite{boys:70} counterpoise method to correct for the basis set
superposition error (BSSE).

In our calculations of the potential surface we kept the C-O bond
lengths fixed and used the O-C-O angle as the bend coordinate, whereas
in the harmonic approximation used in Sec.~\ref{sec:monomer} below the
displacements of the C and O atoms in the CO$_2$ bend coordinate
$\tilde{Q}$ are rectilinear by definition. We tested both alternatives;
some results are displayed in Fig.~S1 of the Supplementary material.
They show that the differences are very small, because the bend
amplitude remains small. Only at very short distances $R$ where the
potential is strongly repulsive, small differences are visible.

The \textit{ab initio} potential was calculated on a grid of 16\,000
symmetry-unique points. This grid contained 25 points for $R$, with step
size $0.25\,a_0$ for $3.5 \leq R \leq 7\,a_0$ and step size $0.5\,a_0$
for $7 \leq R \leq 10\,a_0$, and 4 logarithmically spaced points for $10
\leq R \leq 15\,a_0$. For $\beta'$ we used 8 Gauss-Legendre quadrature
points, ranging from $0$ to $\pi/2$ because of symmetry. For $\gamma'$
we used 16 Gauss-Chebychev equidistant quadrature points between $0$ and
$\pi$, again because of the symmetry. For the bend coordinate
$\tilde{Q}$ we used five points: $\tilde{Q}=0,0.5,1,1.5,2$. For a number
of near-linear geometries with $\beta' = 8.35^\circ$ and the smallest
distance $R=3.5\,a_0$ the He atom is very close to the nearest O atom
and the \textit{ab initio} calculations could not be converged. The
missing data points were provided by exponentially extrapolating the
interaction energies at $R = 4.0$ and $3.75\,a_0$. The potential is
extremely repulsive in this region and the interaction energies are much
higher than the highest collision energy, so these data points do not
play a role in the scattering calculations.

The angular dependence of the 4D potential is represented by the
expansion
\begin{equation}
\label{eq:spher}
 V(\tilde{Q}, R,\beta',\gamma') = \sum_{\lambda=0}^{\lambda_{\rm max}}
 \sum_{m_{\lambda} = 0}^{\lambda}
 v_{\lambda m_{\lambda}} (\tilde{Q},R) \, S_{\lambda m_{\lambda}}(\beta',\gamma')
\end{equation}
in cosine type tesseral harmonics $S_{\lambda m_{\lambda}} =
 \left[ C_{\lambda m_\lambda} +(-1)^{m_\lambda} C_{\lambda -m_\lambda}
                  \right]/\sqrt{2(1+\delta_{m_\lambda 0})}$ with $m_{\lambda}\geq 0$,
which are real-valued linear combinations of the Racah-normalized spherical
harmonics $C_{\lambda m_{\lambda}}(\beta',\gamma')$. The potential is
symmetric with respect to reflection in the $xy$-plane, which implies
that only terms with even values of $\lambda + m_{\lambda}$ occur in the
expansion.

The expansion coefficients $v_{\lambda m_{\lambda}} (\tilde{Q},R)$ were obtained
at each grid point $(\tilde{Q}_k,R_l)$ by numerical integration over $\beta'$
and $\gamma'$ using Gauss-Legendre and Gauss-Chebyshev quadratures,
respectively
\begin{equation}
\label{eq:V_expan}
  v_{\lambda m_{\lambda}} (\tilde{Q}_k,R_l)  = \frac{2\lambda+1}{4\pi}
\sum_{i=1}^{16}
  \sum_{j=1}^{32} w_{i} w'_{j} S_{\lambda m_{\lambda}} (\beta'_{i},\gamma'_{j})
  V(\tilde{Q}_k, R_l,\beta'_{i},\gamma'_{j})
\end{equation}
with weights $w_{i}$ and $w'_{j}$.
The expansion has converged at $\lambda_{\rm max}=15$.
The expansion coefficients $v_{\lambda m_{\lambda}}(\tilde{Q}_k,R_l)$ were
fitted to a fourth-degree polynomial in $\tilde{Q}$
\begin{equation}
\label{eq:V_expan_coeff_fitted_Q2}
v_{\lambda m_{\lambda}} (\tilde{Q},R_l) = \sum_{p=0}^{4}
v_{p,\lambda m_{\lambda}}(R_l) \, \tilde{Q}^p.
\end{equation}
at each grid point $R_l$. It follows furthermore that $m_{\lambda} \leq
p$ and since the potential is invariant under overall rotation about the
$z$-axis, that the sum of $p$ and $m_{\lambda}$ must be even.
Finally, the $R$-dependence of the coefficients $v_{p,\lambda
m_{\lambda}}(R)$ was represented by the Reproducing Kernel Hilbert space
(RKHS) method \cite{ho:96,ho:00}, which uses two parameters: a
smoothness parameter $n$ and a parameter $m$ which ensures that the
potential decays as $1/R^{m+1}$ beyond the largest $R$ value in the
grid. We chose $n=2$ and $m=5$.

\begin {figure}[!t]
\includegraphics*[width=0.80\textwidth]{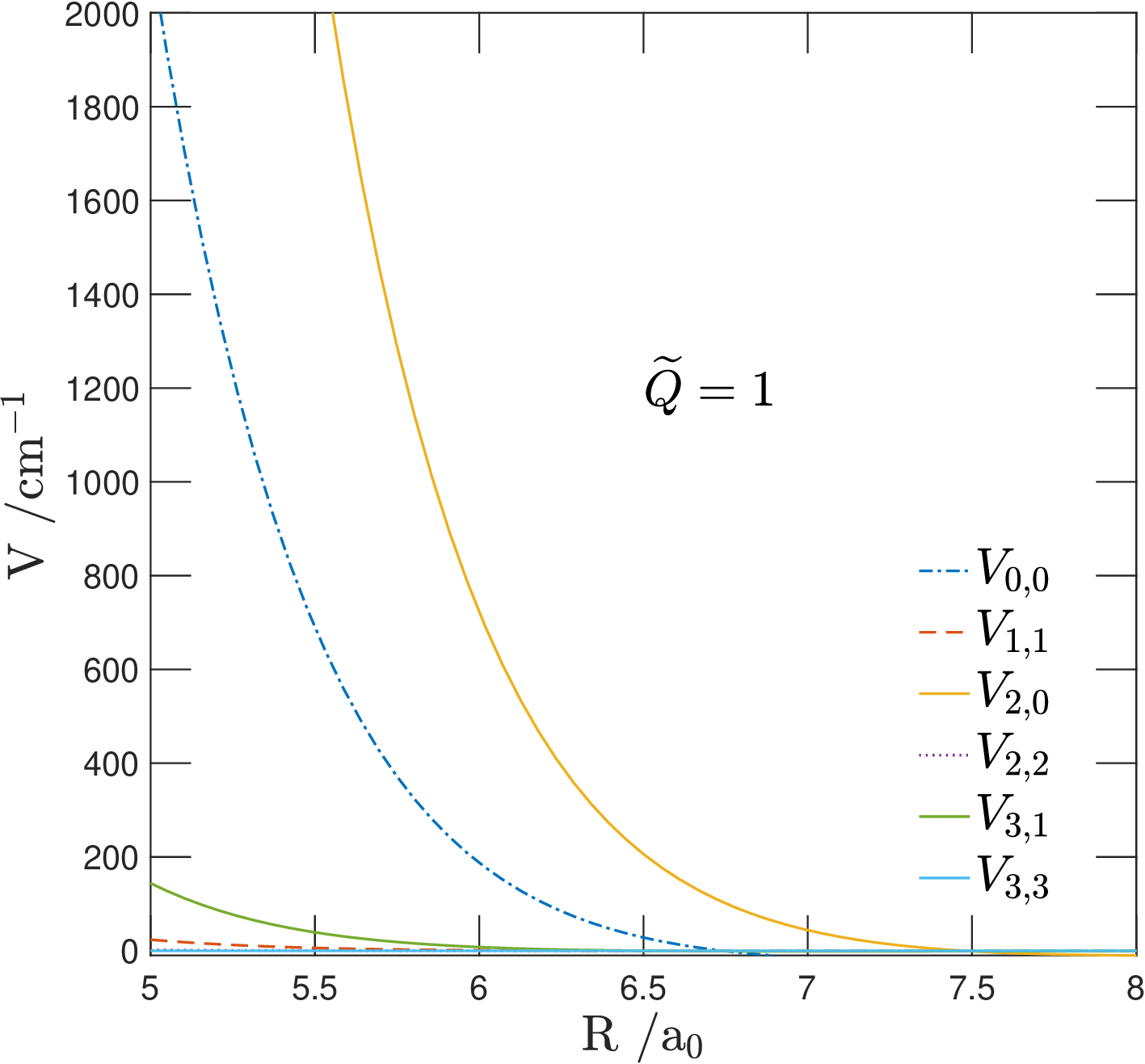}
\caption{
The first few coefficients in the expansion of the potential in
Racah-normalized tesseral harmonics at $\tilde{Q} = 1$.}
\label{fig:IPES_expand}
\end{figure}

In the repulsive short range of the potential at near-linear
configurations extremely high peaks occur at some specific geometries.
These highly repulsive peaks are not physically important because the
system cannot reach these geometries even at high collision energies.
However, they cause a problem with the convergence of the spherical
expansion. In order to avoid this problem we damped the interaction
energies for values larger than $V_0=0.1$\,E$_h$ with the smooth damping
function
\begin{equation}
V_{\rm damped}(\tilde{Q},R,\beta',\gamma') = V_0 + \tanh\left[\zeta
\{V(\tilde{Q},R,\beta',\gamma') - V_0\}\right]/\zeta,
\end{equation}
where $\zeta = 1/V_0$ and $2V_0$ is the maximum value of the
damped potential. The range of $R$ values for which the damping is
effective depends on the angles $\beta'$ and $\gamma'$. At
near-linear structures which $\beta'$ close to 0 and $180^\circ$,
damping was applied for $3.5 \le R \le 4.75\, a_0$ for all $\gamma'$.
For T-shaped structures the interaction energies are less than $V_{0}$ and
damping was not needed.

\begin{figure}[!t]
\includegraphics*[width=0.80\textwidth]{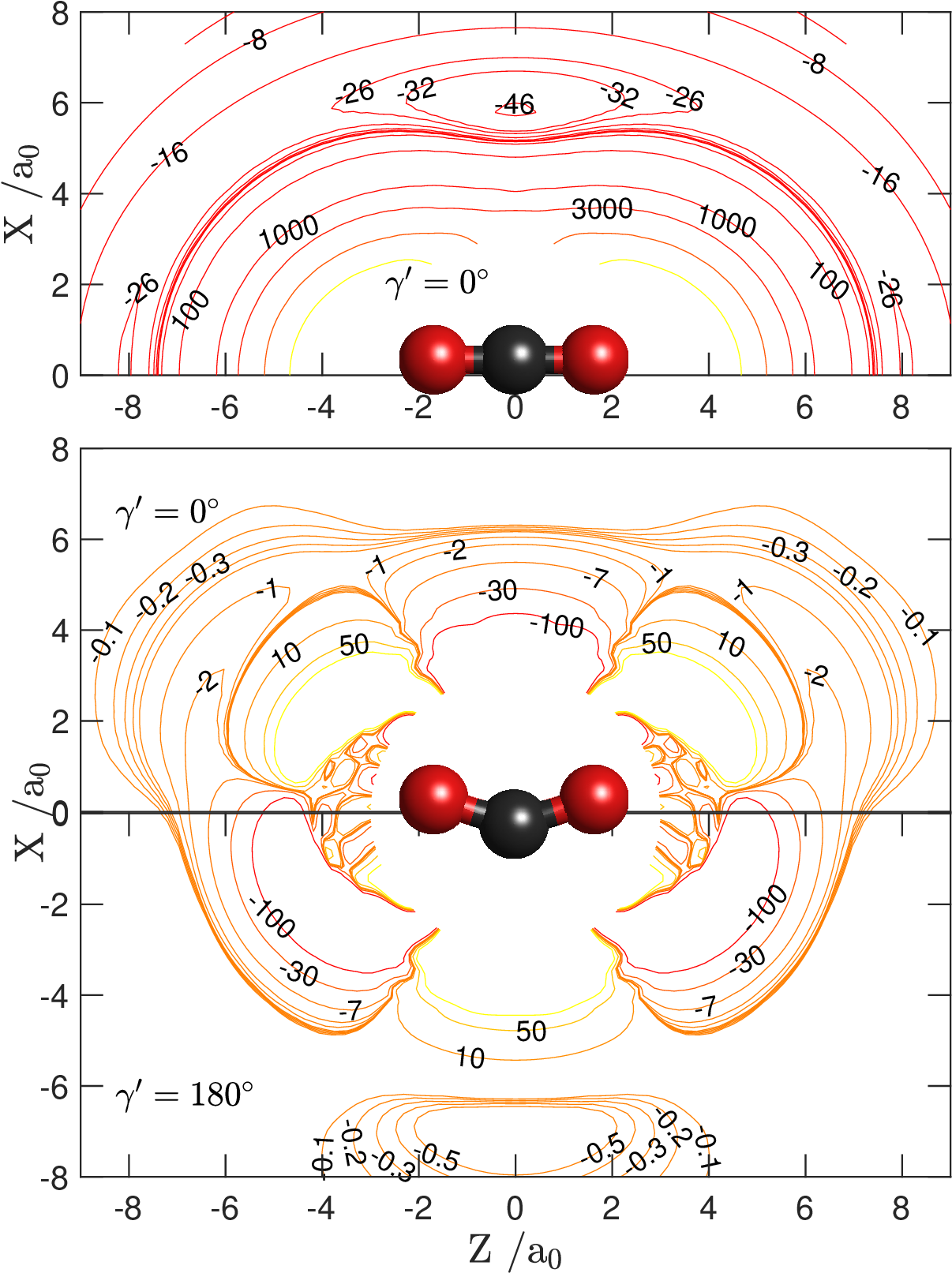}
\caption{Contour plots of the CO$_2$-He potential for $\tilde{Q}=0$ (upper
panel) and the difference between the potential for $\tilde{Q} = 1$ and
$\tilde{Q}=0$ (lower panel). The contours in the lower panel are not drawn in
the region where the repulsive interaction energy is much higher than
the highest collision energy.}
\label{fig:contour}
\end{figure}

The strong anisotropy of the 4D CO$_2$-He potential is obvious already
from the large value of $\lambda_{\rm max}=15$ needed to converge its
spherical expansion. It is further
illustrated by showing the first few expansion coefficients $v_{\lambda
m_{\lambda}}(\tilde{Q},R)$ at $\tilde{Q} = 1$ in
Fig.~\ref{fig:IPES_expand}. The leading anisotropic term $V_{2,0}$ is
larger than the isotropic term $V_{0,0}$, and $V_{3,1}$ is larger than
$V_{1,1}$. At the linear configuration with $\tilde{Q} = 0$, only terms
with even $\lambda$ and $m_{\lambda}= 0$ contribute to the expansion of
the potential. Hence, the terms with odd $\lambda$ and $m_{\lambda}>0$
which are due only to the CO$_2$ bend are smaller.
A view of the CO$_2$-He potential for linear CO$_2$ and the
effect of the bending is shown for planar geometries in
Fig.~\ref{fig:contour}.

Finally we note that the potential $V(\tilde{Q},R,\beta',\gamma')$
calculated with coordinates $\beta',\gamma'$ relative to the MF frame
can simply be expressed relative to the BF frame used in the scattering
calculations by the substitution $\beta' = \beta$ and $\gamma' = \pi -
\gamma$. This yields an additional factor $(-1)^{m_\lambda}$ in the
expansion of the potential in Eq.~(\ref{eq:spher}) when applied to the
coordinates $\beta$ and $\gamma$.

\section{Scattering theory}
\label{sec:theory}
We obtain the cross sections and rate coefficients for rovibrationally
inelastic collisions of CO$_2$ with He by means of scattering
calculations with the numerically exact coupled-channel or
close-coupling (CC) method.
Since the theory for the bend
mode of CO$_2$ is more complicated than described for the two stretch
modes in Refs.~\cite{selim:21,selim:22}, it is outlined below.

\subsection{CO$_2$ monomer Hamiltonian and wave functions}
\label{sec:monomer}
The bend mode of the CO$_2$ monomer is twofold degenerate and the
molecule bends not only in the $xz$-plane but also in the $yz$-plane,
with the coordinates $(Q_x,Q_y)$. The atoms are labeled by the indices
$i=1,2,3$ with $i=1$ and 3 for the two O atoms and $i=2$ for the C atom.
The equilibrium positions in the MF frame are $x_{i}^{e}= y_{i}^{e}=0$,
$z_{1}^{e} = -z_{3}^{e} = 2.196\,a_0$, and $z_{2}^{e} = 0$. We use
rectilinear normal coordinates $Q_{x}$ and $Q_{y}$, which involve the
atomic displacements $\Delta x_{i}$ and $\Delta y_{i}$ from their
equilibrium positions: $\Delta x_{i} = c_{i} Q_x$ and $\Delta y_{i} =
c_{i} Q_y$. The coefficients $c_{1}$ and $c_{3}$ are equal, while
$c_{2}$ is fixed by the condition that the center of mass should not
move.

Our CO$_2$ monomer Hamiltonian is based on Watson's isomorphic
Hamiltonian for the bend mode of a linear triatomic
molecule \cite{watson:70}. In the Cartesian monomer coordinates defined
above and with the use of the harmonic-oscillator rigid-rotor
approximation it reads
\begin{equation}
\label{eq:monham}
  \hat{H}_{\textrm{CO}_2}(Q_x,Q_y) = -\frac{\hbar^2}{2 \mu_{Q}}
  \left( \frac{\partial^{2}}{\partial Q_x^{2}}
       + \frac{\partial^{2}}{\partial Q_y^{2}}  \right)
       + \frac{1}{2} f \left(Q_x^{2}+Q_y^{2} \right)
  + \frac{\hbar^2}{2 I_{0}} \left( \hat{j}_{x}^{2} +  \hat{j}_{y}^{2}
\right),
\end{equation}
where $\mu_Q = 2 m_{\rm C} m_{\rm O}/(m_{\rm C} + 2 m_{\rm O})$ is the reduced mass
associated with the bend vibration, $m_{\rm C}$ and $m_{\rm O}$ are the
masses of the C and O atoms, $f$ is the harmonic force
constant, and $I_{0} = \sum_{i=1}^{3} m_{i} (z_{i}^{e})^2$ is the rigid
rotor moment of inertia for rotation about the $x$- or $y$-axis. The
harmonic frequency is given by $\omega = \sqrt{f/\mu_Q}$. In the sequel
we use dimensionless normal coordinates $\tilde{Q}_{x}=a Q_x$ and
$\tilde{Q}_{y}=a Q_y$ obtained by scaling with $a = \sqrt{\mu_Q \omega/\hbar}$.

Following textbooks \cite{bunker:98} and previous work \cite{banks:87b}
we may re-express this Hamiltonian in
polar coordinates $(\tilde{Q},\gamma)$ with $\tilde{Q}_x = \tilde{Q}
\cos\gamma$ and $\tilde{Q}_y = \tilde{Q} \sin\gamma$
\begin{equation}
\label{eq:monhampolar}
    \hat{H}_{\textrm{CO$_2$}}(\tilde{Q},\gamma) =
     -\frac{\hbar\omega}{2} \left(
      \frac{\partial^2}{\partial \tilde{Q}^{2}}
    + \frac{1}{\tilde{Q}} \frac{\partial}{\partial \tilde{Q}}
    + \frac{\hat{l}_v^2}{\tilde{Q}^2}  \right)
    + \frac{\hbar\omega}{2} \tilde{Q}^{2}
    + \frac{\hbar^2}{2 I_0} \left( \hat{j}^{2} - \hat{j}_{z}^{2}\right),
\end{equation}
where $\tilde{Q}$ is the bend amplitude and $\gamma$ is the angle
between the plane in which CO$_2$ bends and the $xz$ plane. The operator
$\hat{l}_{v} = -i \frac{\partial}{\partial \gamma}$ is the dimensionless
vibrational angular momentum operator with eigenvalues $l_v =
-v,-v+2,\ldots,v-2,v$. The eigenvalues of this harmonic-oscillator
rigid-rotor Hamiltonian are
\begin{equation}
\varepsilon_{v,l_v,j} = (v+1)\hbar\omega + \frac{\hbar^2}{2 I_{0}}
\left[j(j+1) - l_{v}^{2}\right]
\end{equation}
and its eigenfunctions, including rotation with respect to the BF frame,
are
\begin{equation}
\label{eq:moneig}
    \ket{ v, l_v, j,\Omega} = \sqrt{\frac{2j+1}{8\pi^2}} N_{v,l_{v}}
    \tilde{Q}^{|l_{v}|} \,\exp(-\tilde{Q}^2/2) \,
    L_{ (v-|l_{v}|)/2}^{|l_{v}|}(\tilde{Q}) \,
    D_{\Omega, l_{v}}^{(j)}(0,\beta, \gamma)^\ast ,
\end{equation}
where $D_{\Omega,l_v}^{(j)}$ are Wigner $D$-matrices \cite{brink:93}
depending on the rotation angles $(\beta,\gamma)$ of the MF frame
relative to the BF frame, and $\Omega$ is the projection of the monomer
angular momentum $\bmj$ on the BF $z$-axis. Note that in the definition
of the polar coordinates $(\tilde{Q},\gamma)$ above $\gamma$ is the
angle of the plane in which CO$_2$ bends and the functions $\exp(i l_v
\gamma)$ contained in the basis of Eq.~(\ref{eq:moneig}) actually belong
to the vibrational part. The functions
$L_{(v-|l_{v}|)/2}^{|l_{v}|}(\tilde{Q})$ are associated Laguerre
functions in the convention of Abramowitz and Stegun
\cite{abramowitz:64} and the normalization constants

\begin{equation}
\label{eq:normalization}
   N_{v,l_{v}} = \sqrt{2\frac{\left( \frac{ v-|l_{v}|}{2} \right)!}
  { \left( \frac{ v+|l_{v}|}{2} \right)!  }}.
\end{equation}

The functions in Eq.~(\ref{eq:moneig}) may be adapted to the permutation
symmetry $\hat{P}_{13}$ that interchanges the two O nuclei, which has
the following effect
\begin{equation}
\label{eq:P13}
    \hat{P}_{13} \ket{v, l_v, j, \Omega}
        = (-1)^j \ket{v, {-l_v}, j, \Omega}.
\end{equation}
A basis adapted to $\hat{P}_{13}$ with parity $\epsilon$ reads
\begin{equation}
\label{eq:P13adapt}
   \ket{v, \tilde{l}_v,j,\Omega,\epsilon} =
   \frac{1}{\sqrt{2(1+\delta_{l_v,0})}}
   \left[ \ket{v, \tilde{l}_v, j, \Omega}
   + \epsilon (-1)^{j} \ket{v,{-\tilde{l}_v}, j, \Omega}
   \right],
\end{equation}
with $\tilde{l}_v = |l_v|\geq 0$. Since $^{16}$O nuclei are bosons with
spin $I=0$ only functions with $\epsilon = 1$ are allowed, which
implies that for $l_v = 0$ the basis only contains functions with even
$j$.

\subsection{Coupled-channel calculations}
\label{sec:CC}
As outlined in earlier papers \cite{selim:21,selim:22}, our version of
the CC method is based on the CO$_2$-He Hamiltonian in BF coordinates
\begin{equation}
\label{eq:ham}
\hat{H} =  - \frac{\hbar^2}{2 \mu R} \frac{\partial^{2}}{\partial R^{2}} R
 + \hat{H}_{\textrm{CO}_2}(\tilde{Q},\gamma)
 + \frac{\hat{J}^2 + \hat{j}^2 - 2\hat{\bm{j}} \cdot \hat{\bm{J}}}{2 \mu R^{2}}
  + V(\tilde{Q},R,\beta,\gamma),
\end{equation}
where $\mu =
m_{\textrm{CO}_2}m_{\textrm{He}}/(m_{\textrm{CO}_2}+m_{\textrm{He}})$ is
the reduced mass of the complex, $\hat{\bm{j}}$ the CO$_2$ monomer
rotational angular momentum operator, $\hat{\bm{J}}$ the total angular
momentum operator of the complex, and $\hat{J}^2 + \hat{j}^2 -
2\hat{\bm{j}} \cdot \hat{\bm{J}}$ represents the end-over-end angular
momentum operator $L^{2}$ in the BF frame \cite{brocks:83,avoird:94}.
The Hamiltonian $\hat{H}_{\textrm{CO}_2}$ for the CO$_2$ bend
mode is defined in Eq.~(\ref{eq:monhampolar}).

The BF channel basis is
\begin{equation}
    \label{eq:basis}
    \ket{v, \tilde{l}_v, j,\Omega,\epsilon; J, M_J} =
    \sqrt{\frac{2J+1}{4 \pi}}
    \ket{v, \tilde{l}_v,j,\Omega,\epsilon} D_{M_J,\Omega}^{(J)}(\Phi,\Theta,\alpha)^\ast ,
\end{equation}
where $\Omega$ is also the projection of the total angular momentum
$\bmJ$ on the intermolecular axis $\bmR$. The permutation
symmetry-adapted monomer eigenfunctions $\ket{v,
\tilde{l}_v,j,\Omega,\epsilon}$ are defined in Eq.~(\ref{eq:P13adapt}).
The quantum numbers $J$ and $M_J$ are good quantum numbers, while
functions with different $\Omega$ are mixed by the Coriolis coupling
operator $\hat{\bm{j}} \cdot \hat{\bm{J}}$ so that $\Omega$ is an
approximate quantum number.

Another symmetry is the parity of the complex under overall inversion
$E^\ast$. The BF channel basis adapted also to inversion is
\begin{equation}
\label{eq:invadapt}
    \ket{v, \tilde{l}_v, j, \tilde{\Omega};\epsilon, P, J, M_{J}} =
    \frac{1}{\sqrt{2(1+\delta_{\tilde{\Omega},0})}}
    \left[ \ket{v, \tilde{l}_v, j, \tilde{\Omega};\epsilon,J, M_{J}}
     + P\epsilon {(-1)}^{\tilde{l}_v+j+J} \ket{v,\tilde{l}_v,j,{-\tilde{\Omega}};\epsilon, J, M_{J}}
    \right],
\end{equation}
with $\tilde{\Omega} = |\Omega| \geq 0$ and $P = \pm 1$ being the
overall parity. Since $P$ is an exact quantum number the calculations
can be made separately for $P = \pm 1$, which is a considerable
simplification.

In terms of the symmetry-adapted BF channel basis the scattering wave
functions are
\begin{equation}
\label{eq:BF_scattering}
   \Psi^{\epsilon, P,J, M_{J}} = \frac{1}{R} \sum_{v,\tilde{l}_v, j, \tilde{\Omega}}
   \ket{v, \tilde{l}_v, j,\tilde{\Omega};\epsilon, P, J, M_{J}}
   \psi_{v,\tilde{l}_v, j, \tilde{\Omega}}^{\epsilon,P,J, M_{J}}(R).
\end{equation}
The radial functions $\psi_{v,\tilde{l}_v, j,
\tilde{\Omega}}^{\epsilon,P,J, M_{J}}(R)$ can be obtained in the usual
way by solving a set of coupled second order differential equations, the
coupled-channel equations. As in Refs.~\cite{selim:21,selim:22} we do
this with the aid of the renormalized Numerov propagator
\cite{johnson:78,johnson:79}. The required matrix elements of the potential
$V(\tilde{Q},R,\beta,\gamma)$ defined in Sec.~\ref{sec:pot} over
primitive BF channel basis functions $\ket{v,l_v,j,\Omega; J, M_J}$
are given by
\begin{eqnarray}
\label{eq:matel}
V_{v' l'_v j' \Omega'; v j l_v \Omega}^{J M_J} (R) &=&
  \delta_{\Omega'\, \Omega} \; {(-1)}^{\Omega-l'_v}
\frac{1}{\sqrt{2(1+\delta_{m_\lambda 0})}}
 \braopket{v',l'_v}{v_{\lambda m_{\lambda}}(\tilde{Q},R)}{v,l_v}
\nonumber \\
&\times& \threejm{j'}{-\Omega}{\lambda}{0}{j}{\Omega}
\left\{
{(-1)}^{m_{\lambda}}
  \threejm{j'}{-l'_v}{\lambda}{m_{\lambda}}{j}{l_v}
+ \threejm{j'}{-l'_v}{\lambda}{-m_{\lambda}}{j}{l_v}
\right\}.
\end{eqnarray}
The expressions in large round brackets are Wigner $3j$ symbols
\cite{brink:93}. With the symmetry-adapted BF basis of
Eq.~(\ref{eq:invadapt}) one obtains linear combinations of these matrix
elements.

The expansion coefficients $v_{\lambda m_{\lambda}}(\tilde{Q},R)$ are
defined by Eq.~(\ref{eq:spher}) in Sec.~\ref{sec:pot}. Their matrix
elements over the monomer eigenfunctions $\ket{v,l_v}$ defined in
Eq.~(\ref{eq:moneig}) and containing associated Laguerre functions are
calculated by numerical integration over the bend normal coordinate
$\tilde{Q}$ with the aid of a 5-point Gauss-Laguerre quadrature.
Transformation of the matrix elements in Eq.~(\ref{eq:matel}) to the
symmetry-adapted channel basis in Eq.~(\ref{eq:invadapt}) is easy.

The asymptotic boundary conditions to which we need to match the
scattering wave functions at large distance $R$ are defined in terms of
a SF channel basis with partial wave quantum numbers $L$. The monomer
wave functions $\ket{ v, l_v, j,m_j}$ in the SF frame are the same as
those in the BF frame in Eq.~(\ref{eq:moneig}), except that the Euler
angles $(\alpha,\beta,\gamma)$ now define the orientation of the MF
frame relative to the SF frame and the angular momentum component
$\Omega$ on the BF $z$-axis becomes the component $m_j$ on the SF
$z$-axis. The SF monomer basis adapted to $\hat{P}_{13}$ with parity
$\epsilon$ is
\begin{equation}
\label{eq:SFadapt}
   \ket{v, \tilde{l}_v, j, m_j;\epsilon} =
   \frac{1}{\sqrt{2(1+\delta_{l_v,0})}}
   \left[ \ket{v, \tilde{l}_v, j, m_j }
   + \epsilon (-1)^{j} \ket{v,{-\tilde{l}_v}, j,m_j}
   \right]
\end{equation}
and the symmetry adapted SF channel basis is
\begin{equation}
\ket{v,\tilde{l}_v,j,L; \epsilon, P, J, M_{J}}=
\sum_{m_j M_L} \ket{ v, \tilde{l}_v, j,m_j;\epsilon}\, Y_{L M_L}(\Theta,\Phi)\,\braket{j m_j L M_L}{J M_J},
\end{equation}
with $\braket{j m_j L M_L}{J M_J}$ being a Clebsch-Gordan coefficient \cite{brink:93}
and $Y_{L M_L}(\Theta,\Phi)$ a normalized spherical harmonic depending
on the polar angles of $\bmR$ with respect to the SF frame. This basis
is also adapted to overall inversion with parity $P =
\epsilon(-1)^{\tilde{l_v} + L}$.
The primitive BF and SF channel bases are related by the unitary
transformation
\begin{equation}
\label{eq:BFSF}
\ket{v, l_v, j, \Omega; J, M_J}  = \sum_{L}
\ket{v, l_v, j, L; J M_J}  U_{L \Omega}^{J j}
\end{equation}
with
\begin{equation}
\label{eq:U}
  U_{L \Omega}^{J j} = \braket{j\Omega L 0}{J \Omega}
\sqrt{\frac{2L+1}{2J+1}}.
\end{equation}
The unitary transformation between the parity-adapted BF and SF channel
bases then follows from Eqs.~(\ref{eq:P13adapt}), (\ref{eq:invadapt}),
and (\ref{eq:SFadapt}). At the end of the propagation of the BF scattering
functions to large $R$ we transform them to the SF basis
with partial wave quantum numbers $L$ and match them to spherical Bessel
asymptotic boundary conditions to obtain the $S$-matrix. We
then compute the state-to-state scattering cross sections
$\sigma_{{v',\tilde{l}'_v,j' \leftarrow v,\tilde{l}_{v},j}}(E)$ in
the usual way from the $S$-matrix
for a large range of energies $E$ and compute state-to-state rate
coefficients with
\begin{equation}
\label{eq:rate_state_to_state}
  k_{v',\tilde{l}'_v,j' \leftarrow v,\tilde{l}_v,j}(T) =
  \left(\frac{8 k_{B} T}{\pi \mu}\right)^{1/2} \int_{0}^{\infty}
  \sigma_{{v',\tilde{l}'_v,j' \leftarrow v,\tilde{l}_{v},j}}(E)
  \left(\frac{E}{k_B T}\right) \,
  \exp\left(-\frac{E}{k_B T}\right)\, d\left(\frac{E}{k_B T}\right),
\end{equation}
where $k_B$ is the Boltzmann constant. The vibrational quenching rate
from a rovibrational initial state
$(v,\tilde{l}_{v},j)$ to a final vibrational state $(v',\tilde{l}'_v)$
is defined as the sum over all final rotational states $j'$ in $(v',\tilde{l}'_v) $
\begin{equation}
\label{eq:k_vib_quenching}
  k_{v' \tilde{l}'_v \leftarrow v \tilde{l}_{v}, j}(T) =
  \sum_{j'} k_{v' \tilde{l}'_{v} j' \leftarrow v \tilde{l}_{v} j}(T).
\end{equation}
At thermal equilibrium the total vibrational quenching rate
coefficient is computed by Boltzmann averaging over
the thermally populated initial states $j$
in the initial vibrational state $(v,\tilde{l}_{v})$ with energies
$\epsilon_{v \tilde{l}_{v} j}$
\begin{equation}
\label{eq:rate_vib_quenching}
  k_{v' \tilde{l}'_v \leftarrow v \tilde{l}_{v}}(T)
  = \frac{\sum_j (2j+1)\exp(-\epsilon_{v \tilde{l}_{v} j}/k_B T)\,
  k_{v'\tilde{l}'_v \leftarrow v \tilde{l}_{v} j}(T)}
  {\sum_j (2j+1)\exp(-\epsilon_{v \tilde{l}_{v} j}k_B T)}.
\end{equation}

\subsection{Coupled-states approximation}
\label{sec:CSA}
In the coupled-states approximation (CSA) used in the paper by Banks and
Clary \cite{banks:87b} one neglects the Coriolis coupling terms in the
kinetic energy operator that couple BF basis functions with different
$\Omega$. This makes $\Omega$ an exact quantum number, so that the CC
equations can be separated into subsets of equations for each value of
$\Omega$ that are much smaller than the full set of CC equations.
Moreover, the absolute value of $\Omega$ is limited to the smallest
of the initial or final $j$ value in the scattering process, which
reduces the number of subsets to be included. Altogether, this makes CSA
calculations much faster than full CC calculations.

In our version of the CSA method we use the full diagonal part of the BF
kinetic energy $\big[J(J+1) + j(j+1) - 2 \Omega^2\big]/2 \mu R^2$, just
as in Ref.~\cite{banks:87b}. As mentioned in Sec.~\ref{sec:CC}, we
use the renormalized Numerov method \cite{johnson:78,johnson:79} to
obtain the radial scattering wave functions $\psi_{v,\tilde{l}_v,
j,\tilde{\Omega}}^{\epsilon,P,J,M_{J}}(R)$ in
Eq.~(\ref{eq:BF_scattering}). This method propagates the matrices
$\bm{Q}_i$ defined by
\begin{equation}
\bm{\psi}(R_{i-1}) = \bm{Q}_i \bm{\psi}(R_i),
\end{equation}
over a radial grid with points $R_i$ with $i = 1, \ldots, n$. The column
vectors $\bm{\psi}(R_i)$ contain the radial functions
$\psi_{v,\tilde{l}_v, j,\tilde{\Omega}}^{\epsilon,P,J,M_{J}}(R_i)$ at
grid point $R_i$. As mentioned in Sec.~\ref{sec:CC} about the CC method,
the BF Q-matrices at the end of the propagation to large $R$ are
transformed to the SF basis with partial wave quantum numbers $L$ with
the aid of Eqs.~(\ref{eq:BFSF}) and (\ref{eq:U}) and matched to the
proper asymptotic boundary conditions. In the CSA method we obtain
matrices $\bm{Q}_n^\Omega$ for all $\Omega$ values, put them together as
diagonal blocks into a large matrix $\bm{Q}_n$ over all channels,
transform this large Q-matrix from the BF basis to the SF basis with
partial wave quantum numbers $L$ in the same way as in the full CC
method, and use the asymptotic boundary conditions to obtain the full
$S$-matrix from which the CSA cross sections are calculated.

\subsection{Rotational infinite-order sudden approximation}
\label{sec:VCCIOS}
We also calculated vibrational and rovibrational (de-)excitation cross
sections and rate coefficients for CO$_2$-He collisions with the VCC-IOS
approximation. This method was extensively used in the 1980's by Clary
and co-workers \cite{clary:81,clary:81b,clary:82,clary:83,banks:87b}; we
briefly outline the theory for collisions with CO$_2$ in the bend mode.

In the IOS approximation the centrifugal term in the Hamiltonian is replaced by
\mbox{$J(J+1)/2\mu R^2$}, so that $\Omega$ becomes a good quantum number, and the
rotational energy of CO$_2$ is set to zero \cite{pack:74}.  As a result, the
vibrational coupled-channels problem can be solved for fixed orientations of
the molecule. Calculations are done for a set of $J$ values ranging from
$J=0$ to $J=J_\mathrm{max}$. The vibrational wave functions for CO$_2$ in the bend mode
included in Eq.~(\ref{eq:moneig}) are
\begin{equation}
\label{eq:moneigVCC}
    \ket{ v, l_v} = N_{v,l_{v}}
    \tilde{Q}^{|l_{v}|} \,\exp(-\tilde{Q}^2/2) \,
    L_{(v-|l_{v}|)/2}^{|l_{v}|}(\tilde{Q}) \, \frac{\exp(i l_v \gamma)}{\sqrt{2\pi}},
\end{equation}
with ${N}_{v,l_{v}}$ defined in Eq.~(\ref{eq:normalization}). When the
factor $\exp(i l_v \gamma)$ contained in the Wigner matrix $D_{\Omega,
l_{v}}^{(j)}(0,\beta, \gamma)^\ast$ in Eq.~(\ref{eq:moneig}) is removed
from it, the rotational part of the basis $D_{\Omega,
l_{v}}^{(j)}(0,\beta,0)^\ast$ depends only on the angle $\beta$, which
is the angle between the CO$_2$ $z$-axis and the $z$-axis $\bmR$ of the
BF frame. The interaction potential is symmetric with respect to
reflection in the BF $xz$-plane ($\gamma\rightarrow -\gamma$).
Vibrational wave functions that are even or odd for this symmetry
can be obtained by writing $ \exp(i l_{v} \gamma) = \cos (l_{v} \gamma) + i
\sin (l_{v} \gamma)$. The normalized cosine and sine type monomer
eigenfunctions are denoted by $\ket{v,\tilde{l}_v,p}$ with $\tilde{l}_v
= |l_v|$ and $p$ being even ($\rme$) and odd ($\rmo$), respectively. Only the
cosine functions with $p=\rme$ are needed to calculate vibrational cross
sections for transitions involving $v=0,l_v=0$.

The vibrational coupled-channel equations are solved for a grid
of fixed orientations $\beta_i$ of the CO$_2$ monomer $z$-axis in
the BF frame for each value of $J$.
The coupling contains the matrix elements
\begin{equation}
\label{eq:potVCC}
V^{(p)}_{v'\tilde{l}'_v,v \tilde{l}_v}(R,\beta_i) =
    \braopket{v',\tilde{l}'_v, p}
    {V(\tilde{Q},R,\beta_i,\gamma)}{v,\tilde{l}_v, p}
\end{equation}
over the potential $V(\tilde{Q},R,\beta,\gamma)$ defined in
Sec.~\ref{sec:pot}. Integration over the angle $\gamma$, which
determines the plane in which CO$_2$ bends, can be done analytically
since the potential is expanded in cosine type
tesseral harmonics containing $\cos(m_\lambda \gamma)$, see
Eq.~(\ref{eq:spher}), and the basis contains functions
$\cos(\tilde{l}_{v}\gamma)$ or $\sin(\tilde{l}_{v}\gamma)$. The
$\tilde{Q}$-dependent part of the basis contains associated Laguerre
functions and the integration over $\tilde{Q}$ is done numerically with
a Gauss-Laguerre quadrature, as in Sec.~\ref{sec:CC}.

The VCC problem is solved using the same procedure
as outlined for the full CC equations in Sec.~\ref{sec:CC}. This yields
a scattering matrix for each $\beta_i$ and $J$ with elements
$S_{v',\tilde{l}'_v; v,\tilde{l}_v}^{(J,p)}(\beta_i,E)$. The vibrational
(de)excitation cross sections can be calculated from these $S$-matrices
by integration over $\beta$
\begin{equation}
\label{eq:ICS_IOS}
\sigma_{v',\tilde{l}'_v \leftarrow v,\tilde{l}_v}(E) =
   \frac{\pi}{2 k^{2}_v} g_v
\sum_{J,p} (2J+1) \, \int_{0}^{\pi}
\left|S^{(J,p)}_{v',\tilde{l}'_v; v,\tilde{l}_v}(\beta,E)\right|^{2}
\,\sin\beta\,d\beta
\end{equation}
with $k_v^2 = 2\mu (E-\epsilon_v)$, $\epsilon_v$ being the energy of
the initial vibrational state $v$, and $g_{v}$ a degeneracy factor which
depends on $\tilde{l}_{v}$ of the initial state: $g_{v} = 1$ for
$\tilde{l}_v = 0$ and $g_{v} = 1/2$ for $\tilde{l}_{v} > 0$.
In our case $g_v=1/2$ and only $p=\rme$ contributes. We chose
a Gauss-Legendre quadrature grid for the angles $\beta_{i}$, so that the
integral over $\beta$ can be calculated by numerical quadrature.

As explained by Clary in his 1983 paper \cite{clary:83}, VCC-IOS can
also be used to compute rovibrational cross sections. The
theory in Clary's paper is based on the IOS treatment of atom -
symmetric rotor collisions derived by Green \cite{green:79b}. In this
case the rotational states of the molecule are labeled with the quantum
numbers $j,k$ and the IOS angle-dependent $S$-matrix
$\overline{S}^{(J)}_{j' k'; j k}(\beta,\gamma,E)$ depends on two angles
$\beta$ and $\gamma$. These angles determine the orientation of the
molecule in the BF frame and in the IOS scattering calculations they
are fixed.
In the VCC-IOS method applied to CO$_2$-He collisions with
CO$_2$ in the bend mode the matrix $S^{(J,p)}_{v',\tilde{l}'_v;
v,\tilde{l}_v}(\beta,E)$ in Eq.~(\ref{eq:ICS_IOS}) only depends on the
angle $\beta$, however.
In order to make the connection with the theory
in Refs.~\onlinecite{green:79b,clary:83}, we introduce an auxiliary
angle $\chi$, which
is similar to the extra azimuthal angle $\chi'$ defined by Watson
in his derivation of the rovibrational Hamiltonian for linear molecules
\cite{watson:70}. As in the work of Watson, this is a mathematical trick
which does not require any additional physical asumptions. The
vibrational functions $\Psi_{v,l_v}(\tilde{Q},\gamma)$ in
Eq.~(\ref{eq:moneigVCC}) are multiplied by a phase factor $\exp(-i l_v \chi)$
which gives
\begin{equation}
  \label{eq:phase}
  \Psi_{v,l_v}(\tilde{Q},\gamma-\chi) =
  \Psi_{v,l_v}(\tilde{Q},\gamma)\exp(-i l_v \chi)
\end{equation}
and we compensate for this factor by multiplying the pure rotational
wave function $D_{\Omega,l_{v}}^{(j)}(0,\beta,0)^\ast$ in
Eq.~(\ref{eq:moneig}) with $\exp(i l_v \chi)$, so we get
$D_{\Omega,l_{v}}^{(j)}(0,\beta,0)^\ast \exp(i l_v \chi) =
D_{\Omega,l_{v}}^{(j)}(0,\beta,\chi)^\ast$.

In the VCC part of the
calculation, the rotational part of the basis is omitted,
the angle $\beta$ and the
auxiliary angle $\chi$ are fixed, and the $S$-matrix
$\overline{S}^{(J)}_{v',l'_v; v,l_v}(\beta,\chi,E)$ from the VCC
calculations formally becomes a function of two angles, just as in the
IOS treatment of atom - symmetric rotor collisions \cite{green:79b}. The
vibrational angular momentum $l_v$ takes the role of the rotational
angular momentum projection $k$ in the symmetric rotor. In reality one
does not need to vary the angle $\chi$ in the IOS calculations because
it can be shown that this $S$-matrix is
related to the $\beta$-dependent $S$-matrix in Eq.~(\ref{eq:ICS_IOS}) as
\begin{equation}
\label{eq:IOSchi}
\overline{S}^{(J)}_{v',l'_v,\Omega'; v,l_v,\Omega}(\beta,\chi,E) =
S^{(J)}_{v',l'_v; v,l_v}(\beta,E) \exp[i(l'_v-l_v)\chi].
\end{equation}
This requires the $\beta$-dependent VCC $S$-matrix in the complex basis.
The transformation of the VCC $S$-matrices in the symmetry adapted
basis of Eq.~(\ref{eq:ICS_IOS}) to the complex basis is
given by Eqs.~(14) and (15) in Ref.~\cite{clary:83}. For $\tilde{l}'_v=0$
(or $\tilde{l}_v=0$) it is
\begin{equation}
  S^{(J)}_{v',\pm{\tilde{l}'_v};v,\pm{\tilde{l}_v}}(\beta,\chi) =
  \frac{1}{\sqrt{2}} S^{(J,\rme)}_{v',\tilde{l}'_v;v, \tilde{l}_v}(\beta,\chi)
  e^{\pm i (\tilde{l}_v'-\tilde{l}_v)}.
\end{equation}
Then, following Green \cite{green:79b} and Clary \cite{clary:83},
we define rotation-vibration $S$-matrices
\begin{equation}
  \label{eq:IOSrot}
  \overline{S}^{(J)}_{v',l'_v,j',\Omega'; v,l_v,j,\Omega}(E) =
  \delta_{\Omega',\Omega} \braopket{l'_v,j',\Omega}
  {\overline{S}^{(J)}_{v',l'_v; v,l_v}(\beta,\chi,E)}
  {l_v,j,\Omega},
\end{equation}
containing matrix elements of these two-angle dependent $S$-matrices
over the rotational parts of the CO$_2$ monomer basis functions
\begin{equation}
  \label{eq:mondef}
  \ket{l_v, j, \Omega} = \sqrt{\frac{2j+1}{4\pi}}
  D_{\Omega, l_v}^{(j)}(0,\beta, \chi)^\ast.
\end{equation}
The next step is to transform the $S$-matrix to a $\hat{P}_{13}$
adapted basis. In general, each element of the
$S$-matrix in the symmetry adapted basis is a linear combination of four unadapted
$S$-matrix elements (see Eq.~(12) in Ref.~\cite{clary:83}),
but when the final state has $\tilde{l}'_v=0$ and the
initial state has $\tilde{l}_v>0$, as in our case, only two terms remain,
\begin{equation}
  \label{eq:IOSrotsym}
  S^{(J,\epsilon)}_{v',\tilde{l}'_v,j'\Omega;v,\tilde{l}_v,j,\Omega}(E)
   = \frac{1}{\sqrt{2}} \left[
  \overline{S}^{(J)}_{v',\tilde{l}_v',j',\Omega; v,\tilde{l}_v,j,\Omega}(E)
    + \epsilon(-1)^j\,\overline{S}^{(J)}_{v',\tilde{l}'_v,j',\Omega;
v,{-\tilde{l}_v},j,\Omega}(E)
  \right].
\end{equation}
Note that our $\hat{P}_{13}$ symmetry label $\epsilon$ is a good quantum
number, whereas the $\epsilon$ in Clary's paper \cite{clary:83}
differs from ours by a factor $(-1)^j$ for $l_v\neq 0$. For
$l_v=0$ Clary's $\epsilon$ is set to zero.

The rovibrationally inelastic cross sections
can be obtained from the $S$-matrices in Eq.~(\ref{eq:IOSrotsym})
\begin{equation}
\label{eq:ICS_IOSrot}
\sigma_{v',\tilde{l}'_v,j',\epsilon \leftarrow v,\tilde{l}_v,j,\epsilon}(E)
=    \frac{\pi}{k^{2}_v\,(2j+1)} \sum_{J,\Omega} (2J+1) \left|
 S^{(J,\epsilon)}_{v',\tilde{l}'_v,j',\Omega;v,\tilde{l}_v,j,\Omega}(E)
 \right|^{2}.
\end{equation}
The calculation of the matrix elements in Eq.~(\ref{eq:IOSrot}) can
be carried out analytically when the angle-dependent matrices
$\overline{S}^{(J)}_{v',l'_v;v,l_v}(\beta,\chi,E)$ are expanded in
spherical harmonics $Y_{LM_L}$
\begin{equation}
  \overline{S}^{(J)}_{v',l'_v;v,l_v}(\beta,\chi,E) =
  \sqrt{2\pi}\sum_{L,M_L} Y_{L,M_L}(\beta,\chi) S^{(J,L)}_{v',l_v';v,l_v}(E).
\end{equation}
The expansion coefficients are given by
\begin{equation}
\label{eq:ICS_vibrotA}
  S^{(J,L)}_{v',l_v';v,l_v}(E) = \sqrt{2\pi}
  \int_{0}^{\pi} Y_{L,l'_v - l_v}(\beta,0) S^{(J)}_{v',l'_v;v,l_v}(\beta,E)\,
  \sin\beta\, d\beta.
\end{equation}
Only $M_L = l'_v - l_v$ contributes,
because of the phase factor $\exp [i(l'_v - l_v)\chi]$ in Eq.~(\ref{eq:IOSchi}).
The $\sqrt{2\pi}$ normalizes the spherical harmonics with the
azimuthal angle set to zero.
The general expression of the cross sections in terms of the
expansion coefficients is rather complex (see Eq.~(35) of Green \cite{green:79b}),
but again, for our case with $\tilde{l}_v'=0$ and $\tilde{l}_v>0$
it is simpler. It corresponds to Eqs.~(17)-(20) of Ref.~\cite{clary:83}
and in our notation reads
\begin{equation}
\label{eq:ICS_IOSvibrot}
\sigma_{v',\tilde{l}'_v,j' \leftarrow v,\tilde{l}_v,j}(E) =
   \frac{\pi}{2 k^{2}_v} \sum_{L}
   |\braket{L, \tilde{l}_v'-\tilde{l}_v, j, \tilde{l}_v}{j'\tilde{l}_v'}|^2
   \sum_J (2J+1)
   \left| S_{v',\tilde{l}'_v;v,\tilde{l}_v}^{(J,L,\rme)}\right|^{2},
\end{equation}
with the expansion coefficients
$S_{v',\tilde{l}'_v;v,\tilde{l}_v}^{(J,L,\rme)}$ for the symmetrized
$p=\rme$ basis obtained from Eq.~(\ref{eq:ICS_vibrotA}) by writing
$S^{(J,\rme)}_{v',\tilde{l}'_v;v,\tilde{l}_v}(\beta, E)$ instead of
$S^{(J)}_{v',l'_v;v,l_v}(\beta,E)$. For our case with $l'_v=0$ and only
even $j'$ the constant $B$ in Eq.~(19) of Ref.~\onlinecite{clary:83} is
1 for terms with odd values of $j+L$, while it is 0 for even values.

\subsection{Technical details}
\label{sec:details}
In this section we specify the parameters used in our scattering
calculations. The radial grid in the renormalized Numerov propagator
ranges from $R = 3$ to 15\,$a_0$ in 224 equal steps. We started with an
outer $R$ value of 35\,$a_0$, but found out that this value could be
reduced to 15\,$a_0$ without loss of accuracy in the cross sections. We
included bend vibrational states with $v = 0, 1,$ and 2, with
$\tilde{l}_v = 0$, 1, and (0, 2), respectively. Convergence tests were
carried out in which we also included the $v=3$ functions with
$\tilde{l}_v = (1, 3)$ in the channel basis, but the differences in the
cross sections were only about 1\% for low collision energies where
resonances occur and for the highest energies, and less in the
intermediate energy range. So in the final calculations we omitted the
$v=3$ functions. The channel basis contained CO$_2$ monomer rotational
states with a maximum $j$ value of 50 for each $v$. The minimum $j$
value is $\tilde{l}_v$.

The cross sections were calculated for collision energies from 1 to
2000~cm$^{-1}$, with a step size of 0.2~cm$^{-1}$ for $E\le
16$~cm$^{-1}$, 1~cm$^{-1}$ up to 40~cm$^{-1}$, 5~cm$^{-1}$ up to 50~cm
$^{-1}$, 10~cm$^{-1}$ up to 100~cm$^{-1}$, and 100~cm$^{-1}$ up to
2000~cm$^{-1}$. Sharp peaks occur in the cross sections at low energies,
due to resonances, so we had to use a fine energy grid in this region.
The maximum value of the total angular momentum $J$ needed to converge
the cross sections depends on the energy; it was $J_{\rm max} = 20$ for
$E \le 100$~cm$^{-1}$, 35 for $100 < E \le 500$~cm$^{-1}$, 50 for $500 <
E \le 1000$~cm$^{-1}$, 60 for $1000 < E \le 1500$~cm$^{-1}$, and 70 for
$1500 < E \le 2000$~cm$^{-1}$. Not all $J$ values were actually used;
for energies above 16~cm$^{-1}$ $J$ was increased in steps of 4 and the
cross sections were obtained for all $J$'s by cubic spline interpolation
over the available $J$ values. This was allowed because the cross
sections summed over both parities vary smoothly with $J$, except for
the resonances at low energies. The integration over energy in
Eq.~(\ref{eq:rate_state_to_state}) for the rate coefficients was done by
first making a cubic spline interpolation of the cross sections at the
energies for which they were calculated and next applying the
trapezoidal rule on an energy grid with a spacing of 0.2~cm$^{-1}$.
These energies range up to 2000~cm$^{-1}$, which is sufficient to
obtain rate coefficients up to $T=300$\,K. In order to obtain reliable
rates up to $T=500$\,K we extrapolated the exponentially decaying high
energy tail of the integrand in Eq.~(\ref{eq:rate_state_to_state}) by a
simple exponential function $a \exp (-b R)$ with coefficients $a,b$
fitted to the integrand at the highest two energies and extended the
integration up to an energy of 5000~cm$^{-1}$.

In the VCC-IOS calculations the integration over $\beta$ in
Eqs.~(\ref{eq:ICS_IOS}) and (\ref{eq:ICS_vibrotA}) was done numerically
using Gauss-Legendre quadrature. When calculating vibrational
(de)excitation cross sections from Eq.~(\ref{eq:ICS_IOS}) we used only
16 fixed monomer angles $\beta_i$ ranging from 0 to $\pi$, but when
calculating rovibrationally inelastic cross sections from
Eqs.~(\ref{eq:ICS_IOSvibrot}) and (\ref{eq:ICS_vibrotA}) the number of
quadrature points had to be much larger (up to 100) in order to achieve
convergence.

Our computer codes were written in the free and open source script language
{\sc Scilab}, version 6.1.1 \cite{scilab2022} and computations were
done on a cluster of linux servers.

\section{Results and discussion}
\label{sec:results}
\subsection{Cross sections}
Before presenting our results calculated with the potential described in
Sec.~\ref{sec:pot}, let us mention that we also calculated $v=1,
\tilde{l}_v=1 \rightarrow v'=0, \tilde{l}'_v = 0$ quenching cross
sections with the model potential used by Clary \textit{et al.}
\cite{clary:83,banks:87b}, which exponentially depends on the He-O and
He-C distances with parameters based on Hartree-Fock calculations. The
CSA and VCC-IOS cross sections that we calculated with this potential
differ at most by a few percent from the results in
Ref.~\cite{banks:87b}, which confirms the correctness of both their and
our CSA and VCC-IOS programs. The small differences are probably due to
some technical differences between our scattering calculations and those
in Ref.~\cite{banks:87b}, which were not specified in all detail.

\begin{figure}[!ht]
\centering
\includegraphics[scale=0.65]{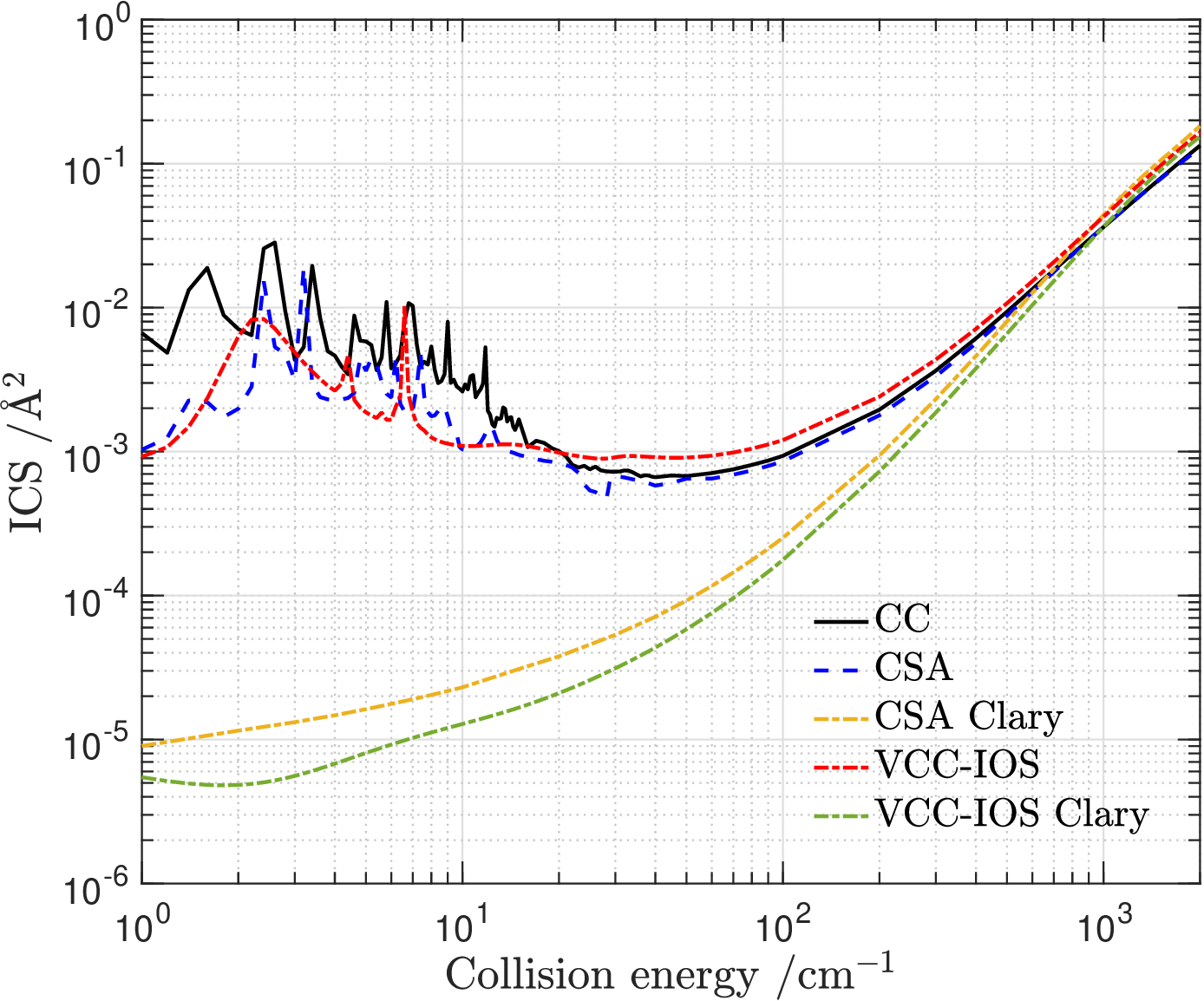}
\caption{Vibrational $v=1, \tilde{l}_v=1 \rightarrow v'=0, \tilde{l}'_v
= 0$ quenching cross sections sections from CC, CSA, and VCC-IOS
calculations with our potential, in comparison with CSA and VCC-IOS
results calculated with the atom-atom model potential of Clary
\textit{et al.} \cite{clary:83,banks:87b}. In the CC and CSA
calculations the initial rotational state has $j=1$ and the cross
sections are summed over all final $j'$ states.}
\label{fig:comptotj1}
\end{figure}

\begin{figure}[!ht]
 \centering
 \includegraphics[scale=0.65]{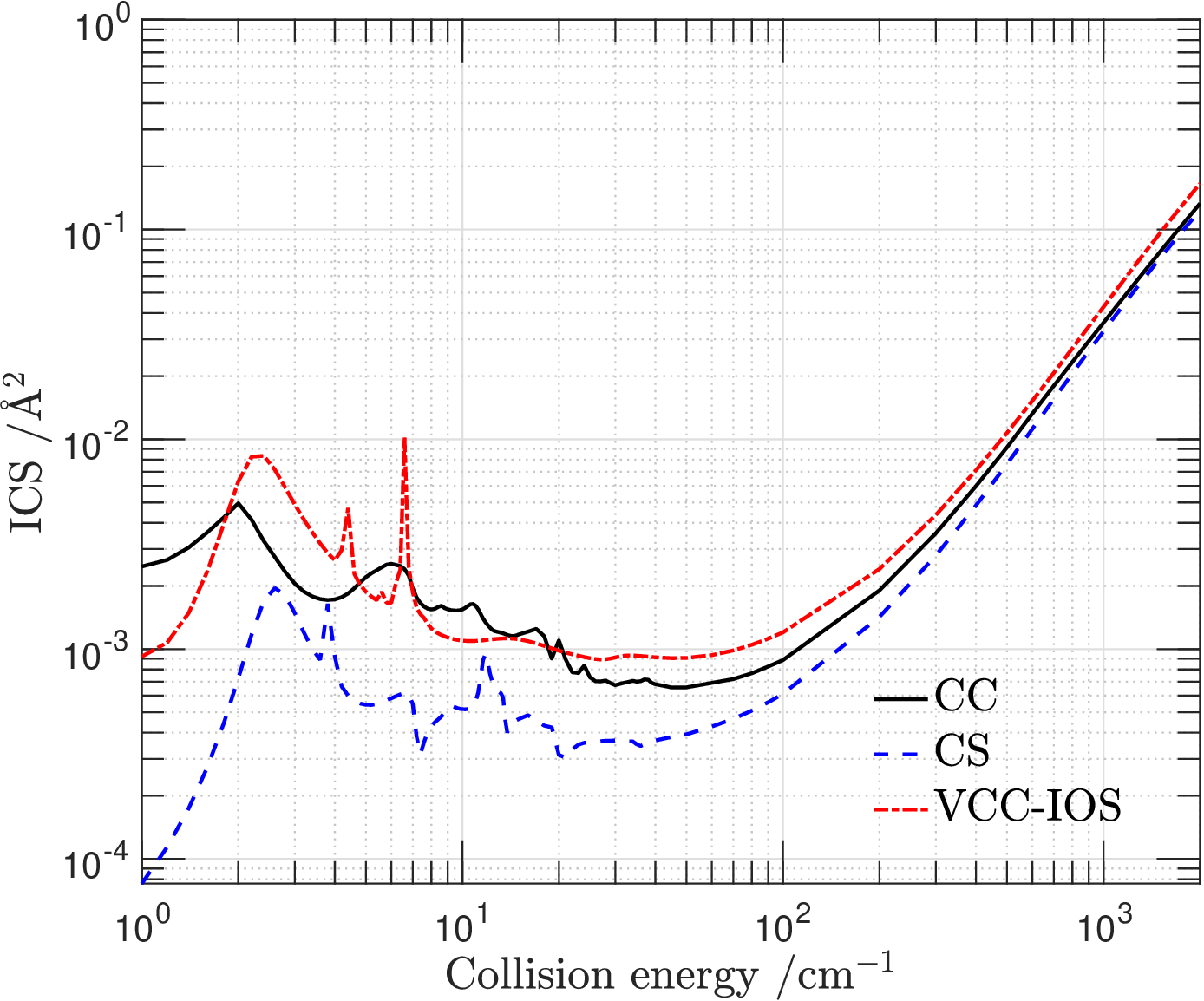}
\caption{Vibrational $v=1, \tilde{l}_v=1 \rightarrow v'=0, \tilde{l}'_v
= 0$ quenching cross sections sections from CC, CSA, and VCC-IOS
calculations with our potential. In the CC and CSA calculations the
initial rotational state has $j=10$, the VCC-IOS results are the same as
in Fig.~\ref{fig:comptotj1}.}
\label{fig:comptotj10}
\end{figure}

Figure~\ref{fig:comptotj1} shows the results of these calculations in
comparison with those obtained with the CC, CSA, and VCC-IOS methods and
our potential in Sec.~\ref{sec:pot}. The results are quite similar for
collision energies above 500~cm$^{-1}$. For the highest energies Clary's
potential yields slightly larger cross sections, which is probably due
to this potential being more strongly repulsive than ours and the
atom-atom model producing a stronger coupling between the $v=1$ and
$v=0$ bend modes of CO$_2$. At energies below 500~cm$^{-1}$ the cross
sections from our potential are much larger, with the difference
increasing to more than two orders of magnitude at collision energies
below 10~cm$^{-1}$. We note incidentally that in Ref.~\cite{banks:87b}
the lowest energy at which the cross sections were calculated was
0.01\,eV, which corresponds to about 80~cm$^{-1}$. Clary's potential
based on SCF calculations completely lacks the attractive dispersion
interactions which are present in our potential and we think that this
explains the much smaller cross sections that it yields at low collision
energies. Another noticeable difference is the presence of scattering
resonances manifested by the peaks in the cross section at energies
below 20~cm$^{-1}$, which are completely missing in the results
calculated with Clary's potential. The reason that the latter potential
does not produce resonances is that it is purely repulsive, so it has no
Van der Waals (VdW) minimum, which implies that it cannot give rise to
bound or quasi-bound states. Our potential based on presently available
computational electronic structure methods is accurate also in the
region of the VdW minimum, and the rate coefficients of rovibrationally
inelastic collisions presented below can also be trusted at low
temperatures.

We see also in Fig.~\ref{fig:comptotj1} that the total vibrational $v=1,
\tilde{l}_v=1 \rightarrow v'=0, \tilde{l}'_v = 0$ quenching cross
section obtained from the approximate CSA and VCC-IOS methods agrees
fairly well with the result from the numerically exact CC method. Only
the subtle resonance structures at low energies are more different, but
these depend very sensitively on the potential and on the scattering
method used. The CSA and CC cross sections in this figure were computed
with the lowest initial $v=1, \tilde{l}_v=1$ rotational state with
$j=1$. Figure~\ref{fig:comptotj10} shows that this holds also for an
initial rotational state with $j=10$, with a resonance structure that is
less pronounced. Comparison of Figs.~\ref{fig:comptotj1} and
\ref{fig:comptotj10} shows, moreover, that the total vibrational
quenching cross section is quite similar for different initial $j$
states.

\begin{figure}[!ht]
\centering
\includegraphics[scale=0.51]{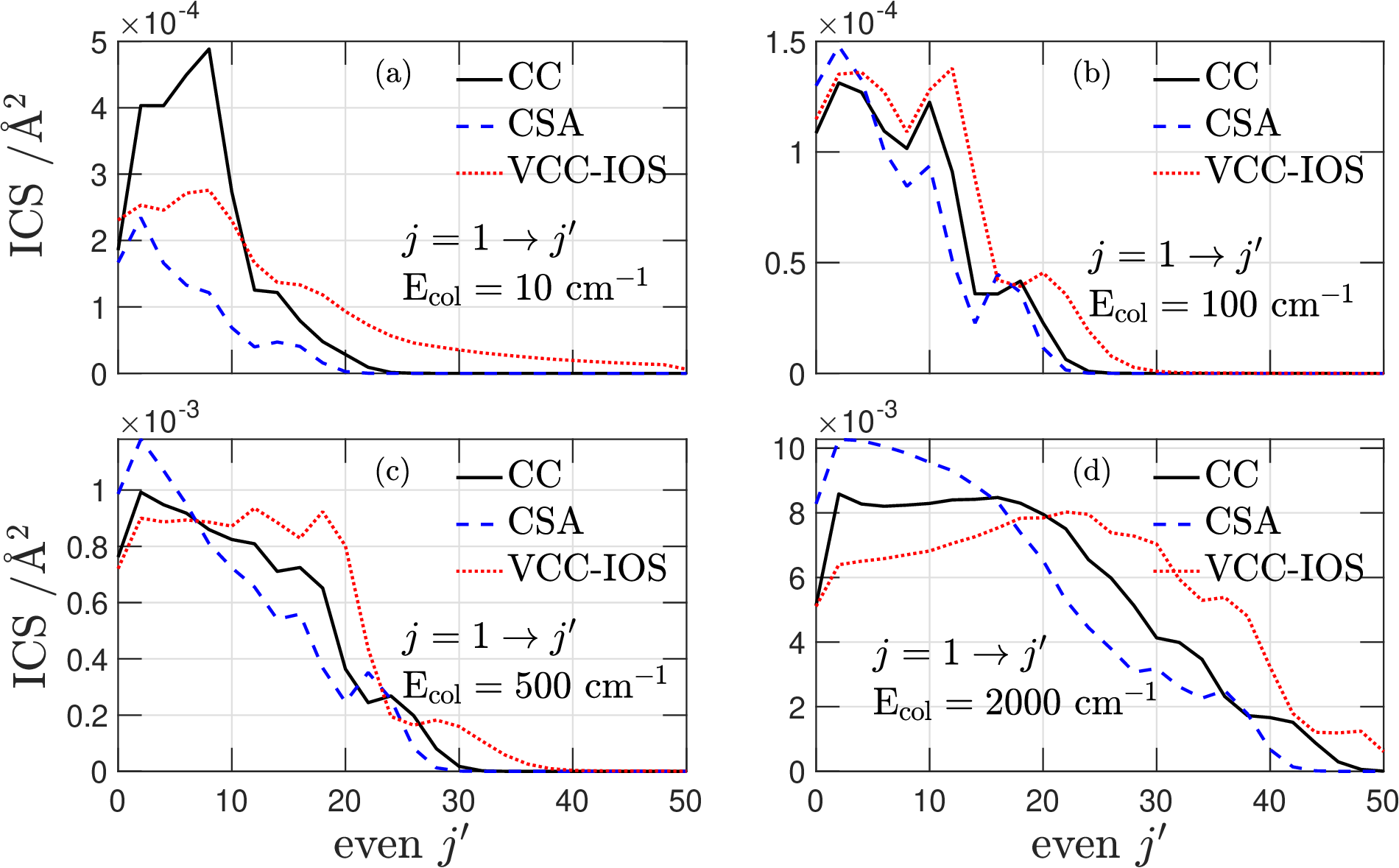}
\caption{Product $j'$ distributions in $v=1,\tilde{l}_v=1,j\rightarrow v'=0,
\tilde{l}'_v=0,j'$ transitions with initial $j=1$ from state-to-state CC, CSA,
and VCC-IOS calculations at different collision energies. Panels (a),
(b), (c), and (d) correspond to collision energies of 10, 100, 500, and
2000~cm$^{-1}$, respectively.}
\label{fig:prodj1}
\end{figure}

\begin{figure}[!ht]
\centering
\includegraphics[scale=0.52]{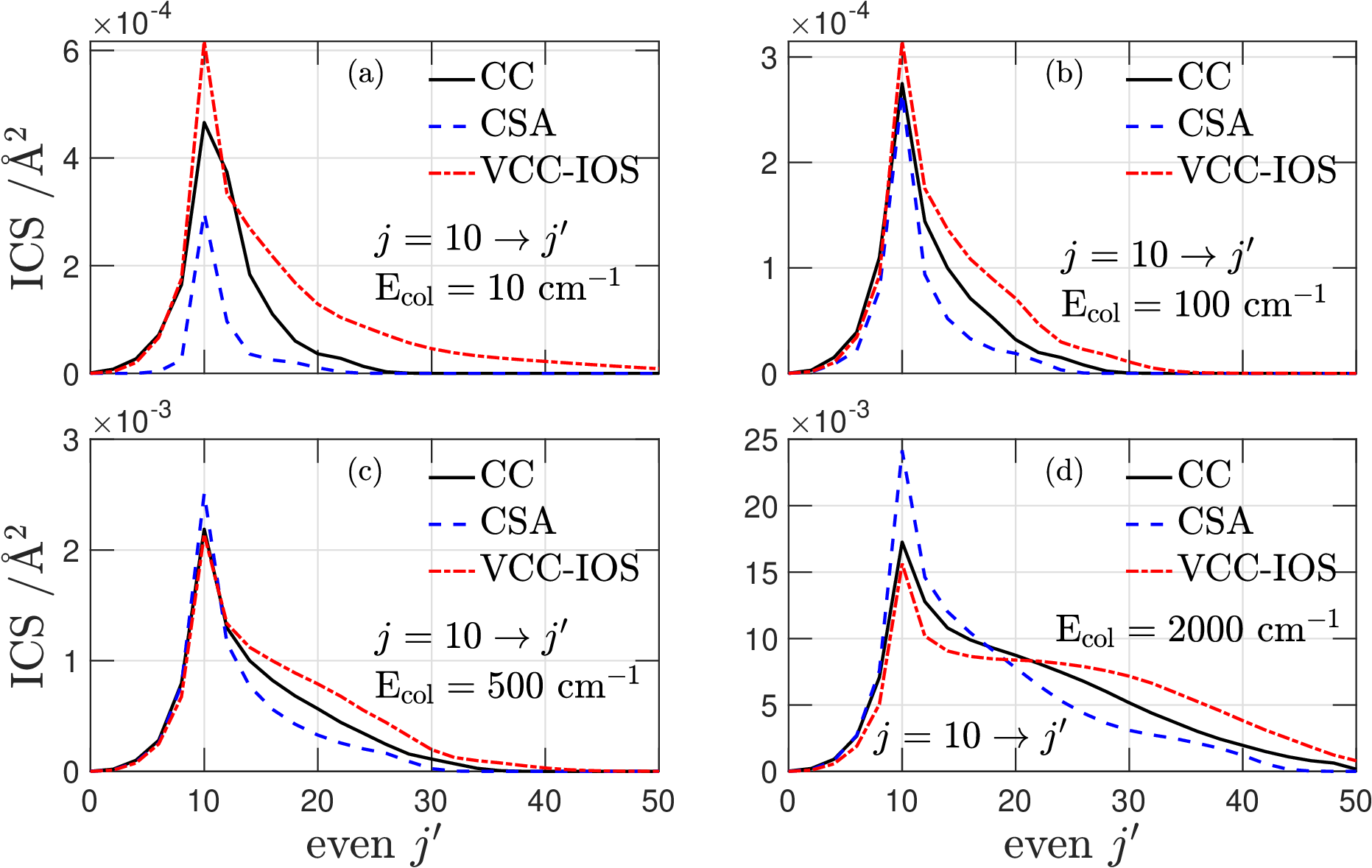}
\caption{Same as Fig.~\ref{fig:prodj1}, with initial rotational state
$j=10$. Panels (a),
(b), (c), and (d) correspond to collision energies of 10, 100, 500, and
2000~cm$^{-1}$, respectively.}
\label{fig:prodj10}
\end{figure}

The CC and CSA methods directly produce rotationally resolved $v=1,
\tilde{l}_v=1, j \rightarrow v'=0, \tilde{l}'_v = 0, j'$ cross sections
and we explained in Sec.~\ref{sec:VCCIOS} how the VCC-IOS method has
been extended \cite{clary:83} to also yield such rotational
state-to-state cross sections. The $j'$ product distributions from the
different methods are displayed for various collision energies in
Figs.~\ref{fig:prodj1} and \ref{fig:prodj10} for initial $j=1$ and
$j=10$, respectively. In Fig.~\ref{fig:prodj1} we observe that for
initial $j=1$ higher and higher $j'$ states are excited when the
collision energy is increased, which is natural of course, and that the
CC, CSA, and VCC-IOS methods show similar trends. The largest
differences occur at the energy of 10~cm$^{-1}$, but this is in the
region where scattering resonances occur. The differences also become
larger at higher collision energies, but they typically stay in the
range of 20 to 50\%, with the CSA method yielding smaller $j'$ values
than CC and the VCC-IOS method yielding larger $j'$. In
Fig.~\ref{fig:prodj10} we observe that for initial $j=10$ there is a
very strong preference for rotationally elastic $j=10 \rightarrow j'=10$
transitions, although also here the $j'$ distribution naturally becomes
wider at higher collision energies. Again, the different methods produce
quite similar results, with the CSA method yielding smaller $j'$ values
than CC and the VCC-IOS method yielding larger $j'$. Also here the
largest deviations occur in the region of the resonances at
10~cm$^{-1}$.

\begin{figure}[!ht] \centering
\includegraphics[scale=0.48]{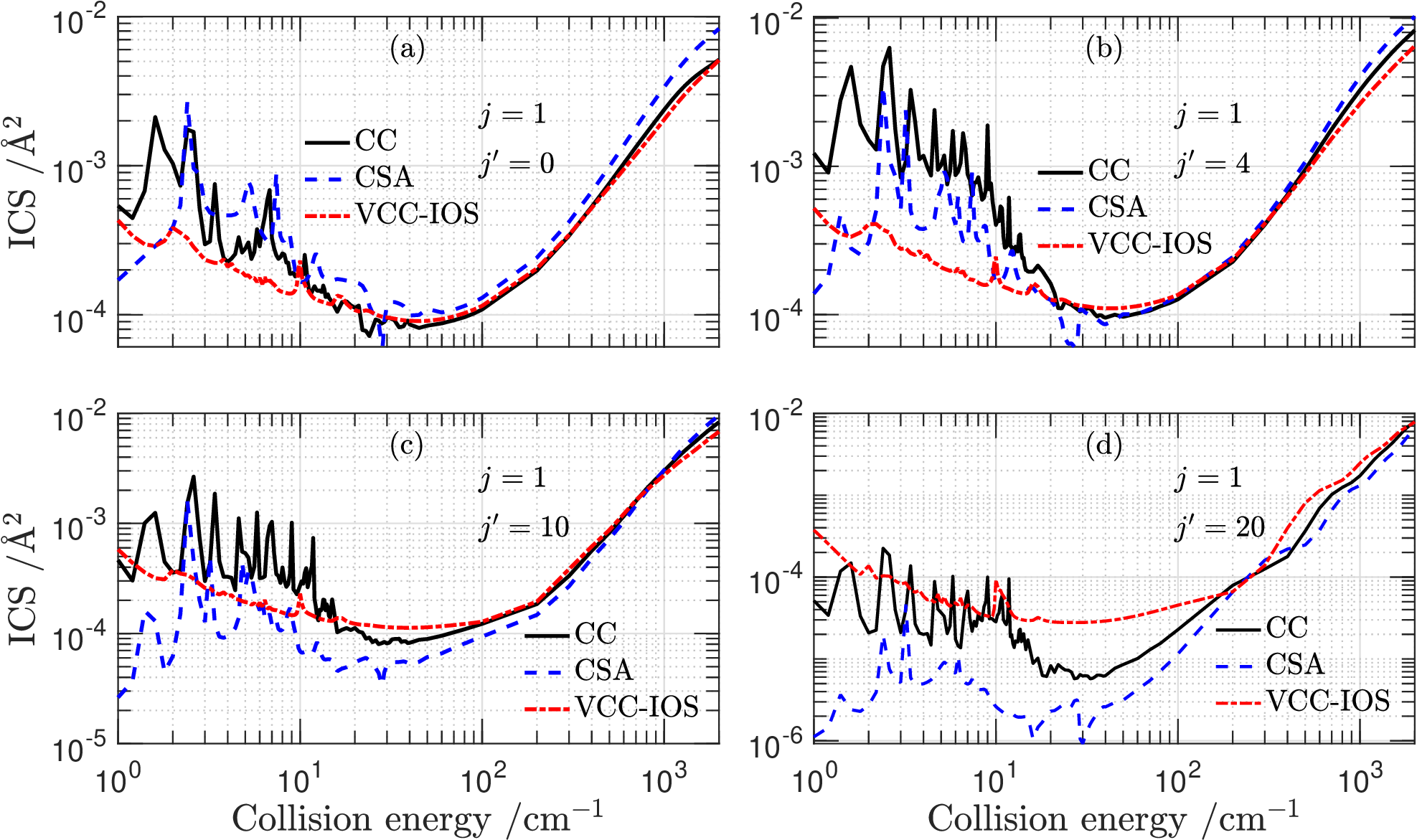}
\caption{State-to-state cross sections as functions of the energy for
$v=1,\tilde{l}_v=1,j\rightarrow v'=0, \tilde{l}'_v=0,j'$ transitions with initial $j=1$
and different $j'$ from state-to-state CC, CSA, and VCC-IOS
calculations. Panels (a), (b), (c), and (d) correspond to
$v'=0$ final states with $j'=0$, 4, 10, and 20, respectively.}
\label{fig:st-to-st_j1}
\end{figure}

\begin{figure}[!ht]
\centering
\includegraphics[scale=0.48]{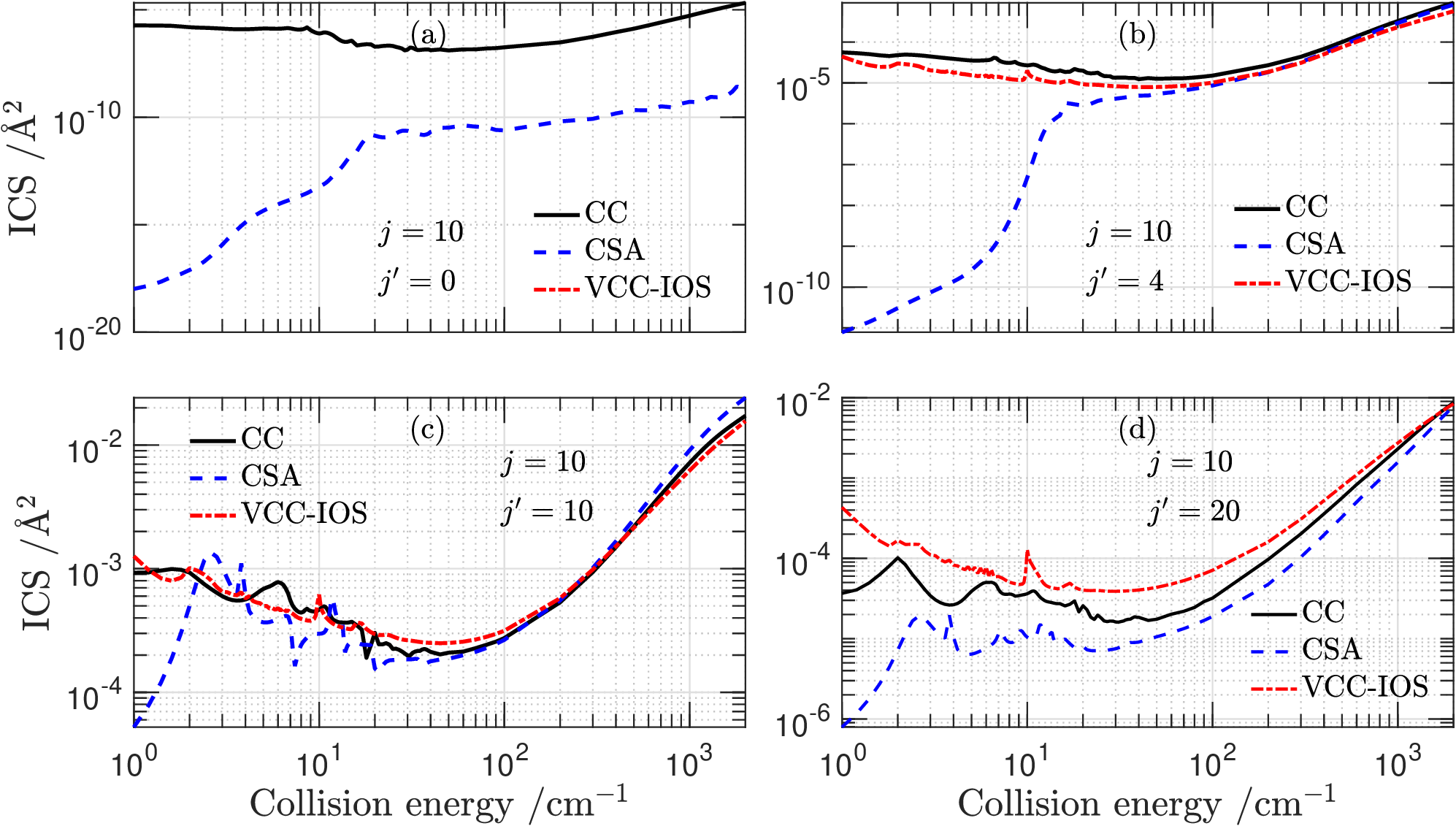}
\caption{Same as Fig.~\ref{fig:st-to-st_j1}, with initial rotation
state $j=10$. Panels (a), (b), (c), and (d) correspond to $v'=0$ final
states with $j'=0$, 4, 10, and 20, respectively.}
\label{fig:st-to-st_j10}
\end{figure}

State-to-state $v=1,\tilde{l}_v=1,j\rightarrow v'=0,\tilde{l}'_v=0,j'$
cross sections from CC, CSA, and VCC-IOS calculations are shown as
functions of the collision energy for different $j'$ values in
Figs.~\ref{fig:st-to-st_j1} and \ref{fig:st-to-st_j10} for initial $j=1$
and $j=10$, respectively. These figures confirm that the cross sections
from the approximate CSA and VCC-IOS methods agree to within 50\% with
those from the full CC method also at the state-to-state level for
energies higher than 30~cm$^{-1}$ and that the deviations become larger
for lower energies where the resonances occur. Also for larger $\Delta j
= |j'-j|$ where the cross sections become smaller the agreement gets
somewhat worse. It is striking that the more approximate VCC-IOS method
yields cross ssections, even at the rotational state-to-state level,
that agree as well with the full CC data as the CSA results, and in some
cases even better.

Apart from the quality of the approximate CSA and VCC-IOS results in
comparison with the full CC data, we may also discuss some general
trends in the cross sections. One can observe in the above figures that
$\Delta j$, the difference between the rotational quantum numbers, is
more significant in determining the magnitude of the cross sections than
the energy gap between the initial $v=1,\tilde{l}_v=1,j$ and final
$v'=0,\tilde{l}'_v=0,j'$ states. We note here that the energy gap
between the lowest vibrationally excited state with
$v=1,\tilde{l}_v=1,j=1$ and the $v'=0,\tilde{l}'_v=0,j'$ state is
smallest for $j'=41$. Generally the cross sections are smallest in the
energy range between 30 and 100~cm$^{-1}$ and increase by at least two
orders of magnitude when the energy is raised to 2000~cm$^{-1}$. This
increase is larger when $\Delta j$ is larger.

A feature in Figs.~\ref{fig:st-to-st_j1} and \ref{fig:st-to-st_j10} that
is particularly striking is that the cross sections from CSA
calculations become very small for collision energies below 2~cm$^{-1}$.
This may be explained as follows. In the BF coordinates on which the CSA
method is based the centrifugal barrier is represented by a term that
includes diagonal and off-diagonal Coriolis couplings, see
Eq.~(\ref{eq:ham}). The latter are omitted in CSA, which effectively
increases the height of the centrifugal barrier. This implies that at
low energies the colliding CO$_2$ molecule and He atom are more strongly
prevented from getting closer, where the potential coupling is larger
and the vibrational transitions occur. We confirmed this explanation by
CSA calculations in which we lowered the centrifugal barrier by adding a
term to the diagonal angular kinetic energy that more or less
compensates for the omission of the off-diagonal terms. Indeed, we found
that this brings the CSA cross sections at low energies substially
closer to the CC results. In Fig.~\ref{fig:st-to-st_j10}(a) the CSA
cross sections are smaller than those from CC even at higher energies,
but also the CC cross sections are extremely small in that case.

Another interesting observation regards the partial cross sections from
different total angular momenta $J$ and different parities. Instead of
the total parity $P$ with respect to overall inversion, we consider the
spectroscopic parity $P(-1)^J$. States with even and odd spectroscopic
parity are conventionally labeled $e$ and $f$, respectively. The partial
cross sections for different $J$ values summed over both parities are
displayed in Fig.~S2. One observes, quite naturally, that for higher
collision energies the contributions from higher $J$'s become more
important. And that the maximum values of $J=20$, 50, and 70 used in
different energy ranges produce well converged cross sections. Since the
initial monomer angular momentum is $j=1$ in this example, the total
angular momentum $J$ is nearly equal to the orbital angular momentum
$L$, see Sec.~\ref{sec:CC}, which is related to the impact parameter $b$
as $L=\mu b v$. From the data shown in Fig.~S2 it follows then that the
impact parameters are $b=3.24$, 2.05, and 2.08\,$a_0$ for collision
energies of 100, 1000, and 2000~cm$^{-1}$. These impact parameters are
small, especially at higher energies, which indicates that
(ro)vibrational transitions mostly take place for nearly head-on
collisions and happen in the region where the potential is strongly
repulsive and the coupling potential is relatively large, see
Fig.~\ref{fig:contour}.

\begin{figure}[!ht]
\centering
\includegraphics[scale=0.48]{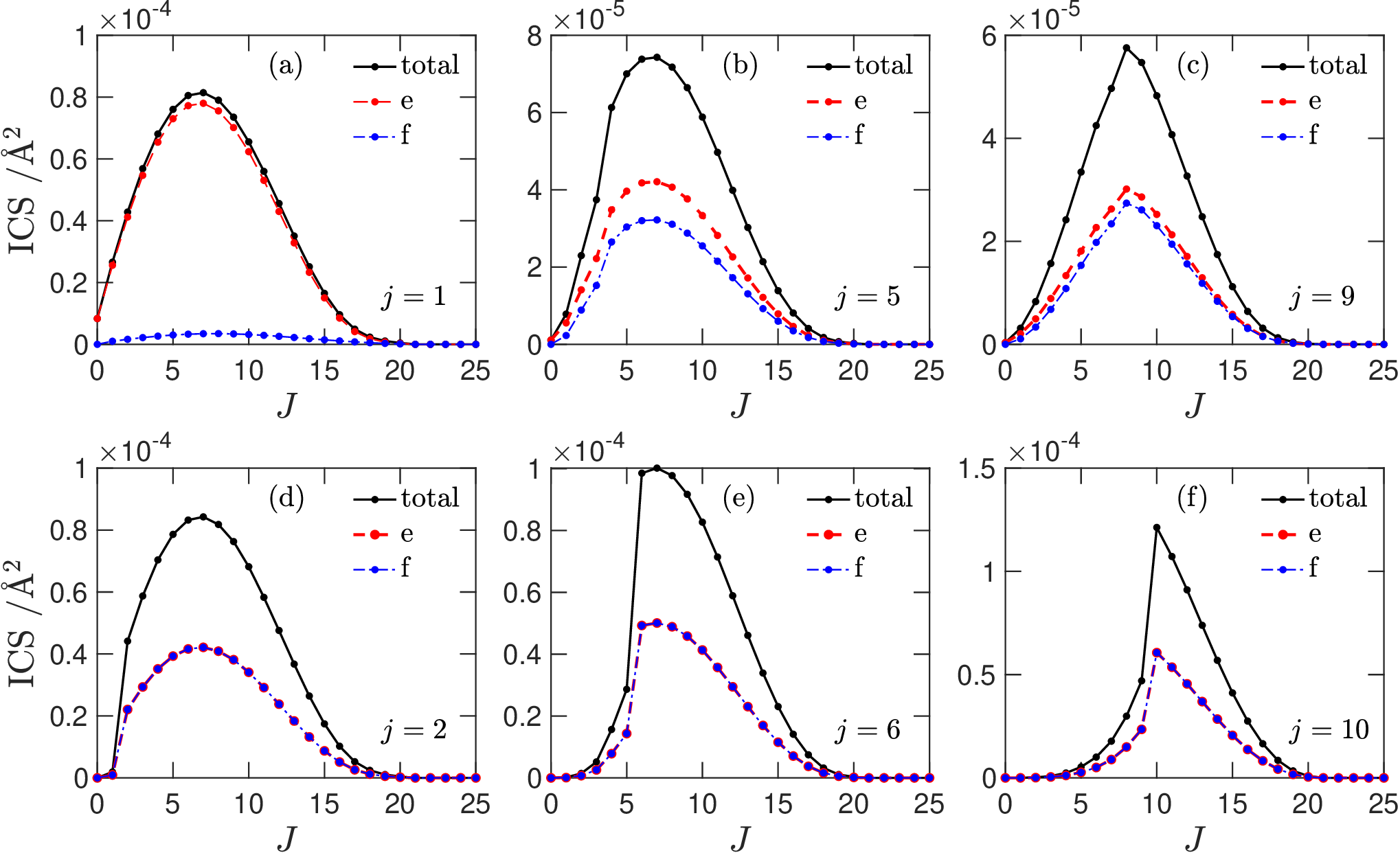}
\caption{Partial wave distributions of the $v=1, \tilde{l}_v=1, j$ to
$v'=0, \tilde{l}'_v=0$ cross sections computed at the CC level at
collision energy $E=100$~cm$^{-1}$ summed over all final $j'$ values and
resolved by spectroscopic parity $e$ or $f$. Panels (a), (b), and (c)
correspond to the odd initial $j$ values 1, 5, and 9, panels (d), (e),
and (f) to the even $j$ values 2, 6, and 10. The closed black curves are
the total partial cross sections, the dashed red and dot-dashed blue
curves are the contributions from parities $e$ and $f$, respectively.
Note that for odd initial $j$ the contributions from parity $e$ are
dominant, especially for $j=1$, while for even $j$ the contributions
from parities $e$ and $f$ are nearly equal.}
\label{fig:partial_wave_parity}
\end{figure}

The relative contributions from the scattering states with spectroscopic
parities $e$ and $f$ to the total cross sections are less obvious, see
Fig.~\ref{fig:partial_wave_parity}. For initial $j=1$ the contributions
from the $e$ states are very dominant, and also for higher odd $j$
values the $e$ states contribute more to the cross sections than the $f$
states. For even initial $j$ values, on the other hand, the
contributions from the $e$ and $f$ states are practically equal. This
can be explained by considering the role of the approximate quantum
number $\Omega$ in the BF formalism explained in Sec.~\ref{sec:CC}. The
absolute value of $\Omega$ is limited by the initial $j$ value, the
final $j'$ value, and the total angular momentum $J$. When analyzing the
state-to-state cross sections, we found that the largest allowed
absolute $\Omega$ values yield the largest contributions. It follows
from Eq.~(\ref{eq:P13adapt}) that for the ground state with
$v'=\tilde{l}'_v=0$ only states with even $j'$ are allowed, and from
Eq.~(\ref{eq:invadapt}) that for $j'=\Omega = 0$ only states with parity
$e$ occur. Furthermore, it follows from the second and third $3j$
symbols in Eq.~(\ref{eq:matel}) with $l_v = 1$ and $l'_v = 0$ that
$m_\lambda$ must be odd, from the reflection symmetry of the potential
in Sec.~\ref{sec:pot} that also $\lambda$ must then be odd, and from the
first $3j$ symbol in Eq.~(\ref{eq:matel}) that for $\Omega=0$ the sum of
$j$ and $j'$ must be odd as well. We already derived that $j'$ must be
even, so only for odd initial $j$ values the $\Omega = 0$ states
contribute to the cross sections. Because these $\Omega = 0$ states are
purely of parity $e$, this explains the observed dominance of the $e$
states over the $f$ states in the cross sections with odd initial $j$.
For the initial states with even $j$ the cross sections do not contain
this $\Omega=0$ component with pure $e$ parity and therefore the $e$ and
$f$ parity contributions are equal. The dominance of the $e$ states over
the $f$ states for odd initial $j$ values is most pronounced for initial
$j=1$ and decreases with increasing $j$. This follows because for the
initial $v=\tilde{l}_v=j=1$ state $|\Omega|$ is limited to 0 and 1, and
the restriction to even $j'$ implies that only the $\Omega=0$ state with
parity $e$ contributes to the cross section through the potential, see
Eq.~(\ref{eq:matel}). For larger $j$ and $j'$ also $\Omega \ne 0$ states
contribute directly to the cross sections through the potential and the
relative contribution from the $\Omega=0$ states becomes less important.

Another feature that can be observed in
Figs.~\ref{fig:partial_wave_parity}(d), (e), and (f) where the cross
sections do not contain the $\Omega=0$ contribution is that for total
$J$ smaller than $j$ and $j'$, where the maximum $|\Omega|$ is limited
by $J$, the partial cross sections remain small and rise steeply when
$J$ becomes equal to the minimum of $j$ and $j'$. This confirms that,
indeed, functions with $|\Omega|$ equal to min$(j,j')$ yield the largest
contributions to the cross sections. Since $\Omega$ is the projection of
the CO$_2$ angular momentum $\bmj$ on the intermolecular axis $\bmR$
this suggests that for rovibrational transitions involving the CO$_2$
bend mode sideways collisions are most effective. Figure~S3 in the
Supplementary material shows the $e/f$ parity-resolved state-to-state
cross sections for initial states with $j=9$ and $j=10$ and final $j'$
values ranging up to 25. This figure shows that our findings regarding
the different parity contributions to the cross sections for odd and
even initial $j$ values also hold for the state-to-state cross sections:
the $e$ parity contributions dominate over those of $f$ parity
for odd $j=9$, especially for low final $j'$, while the $e$ and $f$
contributions are equal for even $j=10$. It also shows that the $j'$
dependence of the state-to-state cross sections is similar to the total
$J$ dependence of the partial cross sections in
Fig.~\ref{fig:partial_wave_parity}, which follows from the limitation of
the maximum $|\Omega|$ by both $j'$ and $J$.

\subsection{Rate coefficients}
\begin{figure}[!t]
\centering
\includegraphics[scale=0.49]{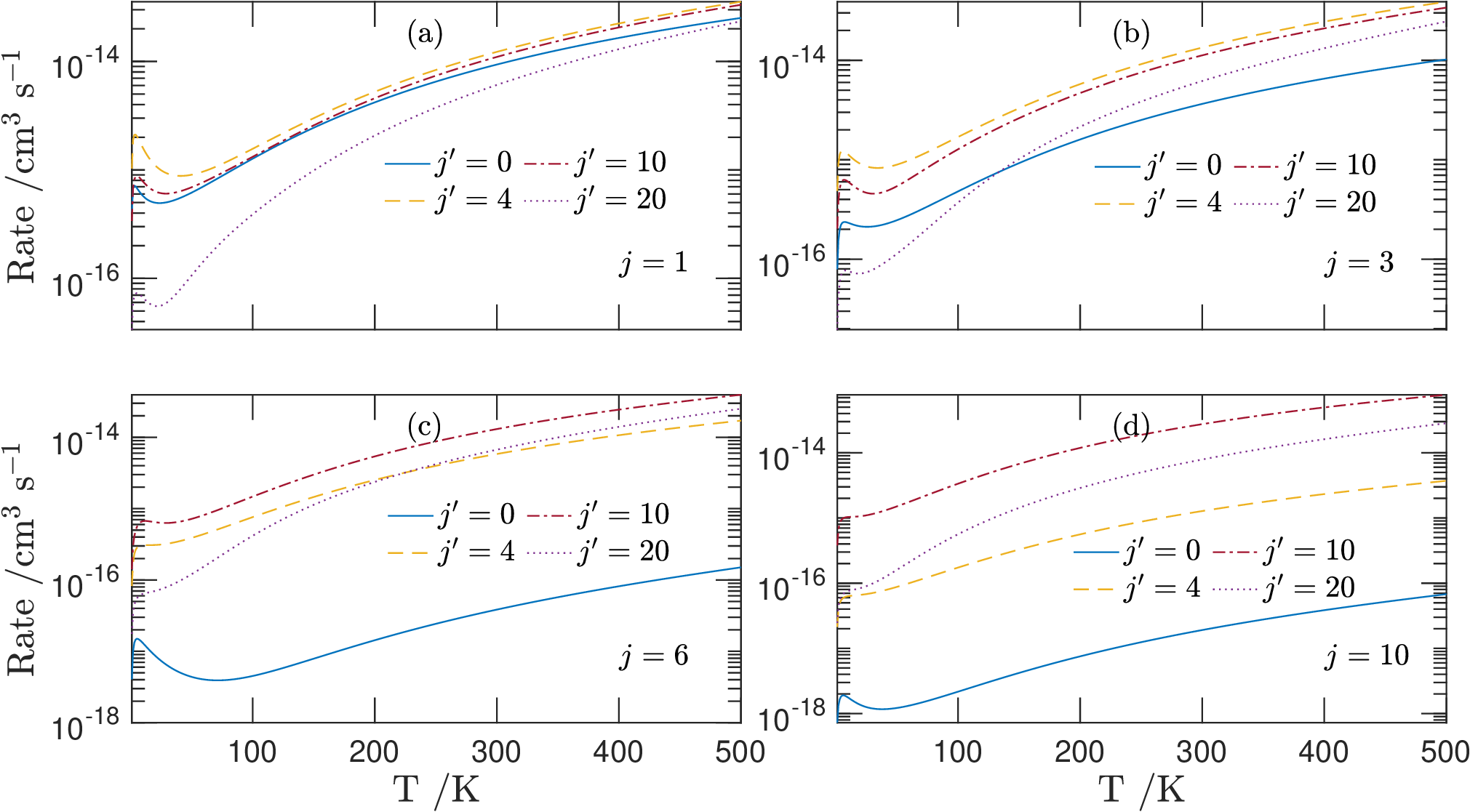}
\caption{State-to-state transition rate coefficients from
CC calculations for different initial $j$ and final $j'$ values.
Panels (a), (b), (c), and (d) correspond to $v=1$ initial
states with $j=1$, 3, 6, and 10, respectively.}
\label{fig:rates_sts}
\end{figure}

In Fig.~\ref{fig:rates_sts} we display some state-to-state
$v=1,\tilde{l}_v=1,j\rightarrow v'=0, \tilde{l}'_v=0,j'$ rovibrational
transition rates from CC calculations for different initial $j$ and
final $j'$ values as functions of the temperature. Such rovibrational
state-to-state rates are the data needed in radiative transfermodels.
For lower initial $j$ the rates of transitions to different $j'$ states
differ by at most one order of magnitude at low temperature and become
more similar with increasing temperature. For higher initial $j$ one
clearly observes that transitions to $j'=j$ are the strongest and that
especially transitions to $j'$ smaller than $j$ are weak. More
generally, these features indicate that transitions with small $\Delta j
= |j'-j|$ are favored and that transitions with $j'>j$ are stronger than
those with $j'<j$. The latter propensity can be explained by the energy
gap law: the energy difference between the initial $v=1,j$ and final
$v'=0,j'$ states is smaller when $j'>j$. We will include a more complete
set of rovibrational transition rates in the LAMDA data base
\cite{lamda:05}.

Figures~S4 and S5 in the Supplementary material show the state-to-state
$v=1,\tilde{l}_v=1,j\rightarrow v'=0, \tilde{l}'_v=0,j'$ rovibrational
transition rates for initial $j=1$ and 10 and different final $j'$
values from CC, CSA, and VCC-IOS calculations. Both the CSA and VCC-IOS
rates agree to within about 50\% with the CC results, just as the
corresponding cross sections, except when $j'$ is smaller than $j$ and
the rates become insignificant.

The rates increase monotonically with the temperature, except for
transitions involving low $j$ or low $j'$ states for which the rates
have minima around 50\,K. The relatively large rates at low temperature
for these transitions are due to contributions from scattering
resonances, as shown in Fig.~\ref{fig:st-to-st_j1}. This figure shows
the strongest scattering resonances in the cross sections from CC
calculations, weaker ones in the CSA results, and practically none in
the VCC-IOS cross sections, which clearly explains why the minima in the
rate coefficients from the CC, CSA and VCC-IOS methods in Figs.~S6 and
S7 of the Supplementary material are strongest for CC, weaker for CSA,
and absent for VCC-IOS.

In Figs.~S6 and S7 of the Supplementary material we compare the total
vibrational quenching rates, i.e., the rovibrational transition
rates summed over all final $j'$ values, calculated with the different
methods, with initial $j=1$ and 10 in the CC and CSA methods. No initial
$j$ value is defined in the VCC-IOS method, so the VCC-IOS quenching
rates are identical in the two figures. Also the CC and CSA quenching
rates are quite similar for $j=1$ and $j=10$, except at the lowest
temperatures where the minima are more pronounced for initial $j=1$, due
to the stronger resonance contributions at low $j$. We observe that the
CSA method somewhat underestimates the rates from CC calculations, while
the VCC-IOS method yields slightly higher rates. The differences are
quite small, however, on the order of 20\% or less. These differences
between the approximate CSA and VCC-IOS methods and the CC method are
smaller than in many of the state-to-state rate coefficients shown in
Figs.~S4 and S5, which is because the smaller state-to-state rates are
more sensitive to the approximations made than the most significant
ones. The larger total vibrational quenching rates from VCC-IOS and
smaller rates from CSA, as compared to the CC results, are due to the
contributions from higher final $j'$ values such as shown in Figs.~S4(d)
and S5(d). We already concluded from Figs.~\ref{fig:prodj1} and
\ref{fig:prodj10} that these are favored by VCC-IOS relative to CC and
disfavored by CSA.

\begin{figure}[!ht]
\centering
\includegraphics[scale=0.65]{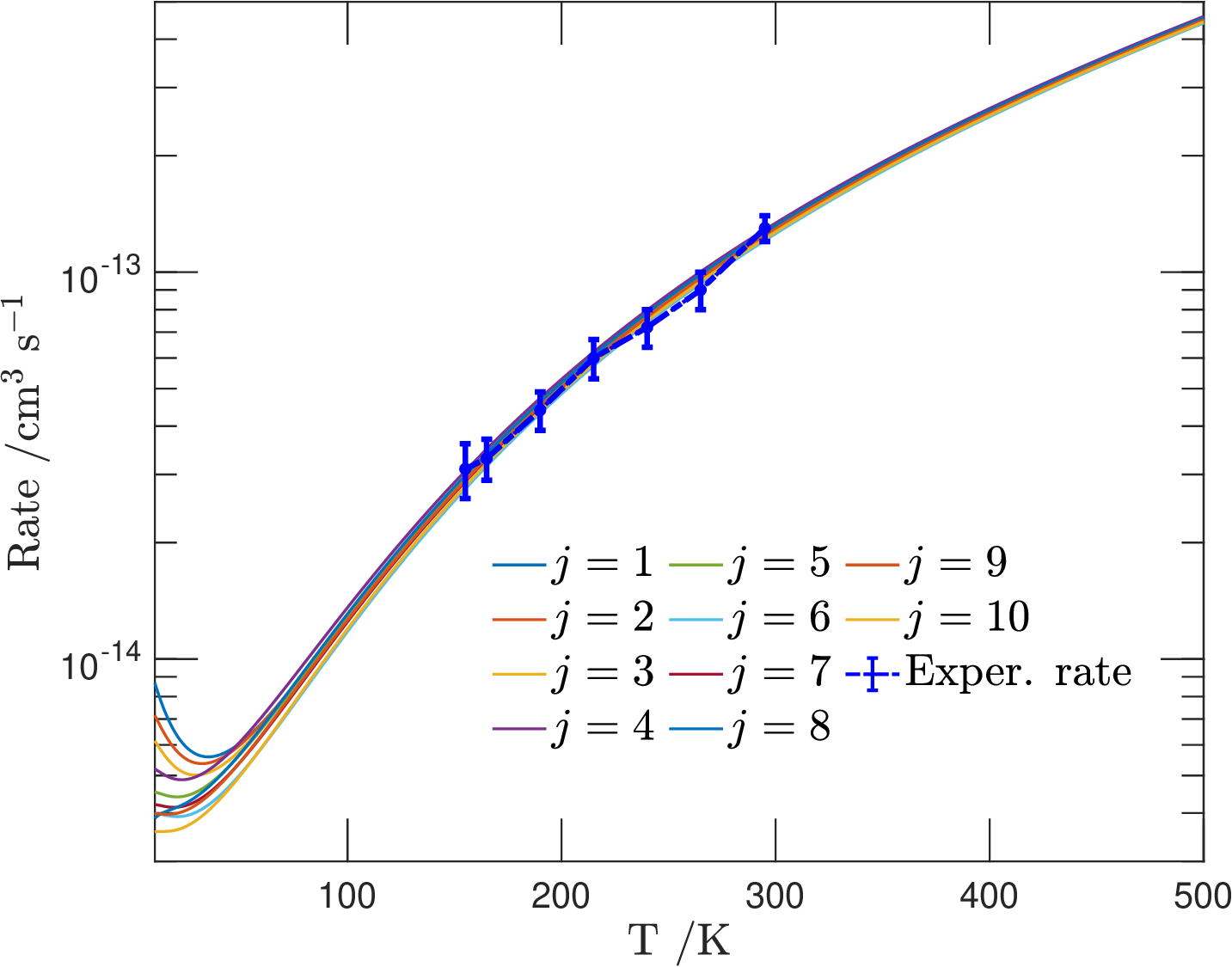}
\caption{Vibrational quenching rates from CC calculations for $v=1,
\tilde{l}_v=1,j\rightarrow v'=0, \tilde{l}'_v=0$ transitions with
different initial $j$ values ranging from 1 to 10 compared with
experimental results \cite{siddles:94}.}
\label{fig:rates_tot}
\end{figure}

Figure~\ref{fig:rates_tot} shows the total vibrational quenching
rates for initial $j$ values ranging from 1 to 10. It shows even more
clearly that the total vibrational quenching rate hardly depends of the
initial $j$, except at temperatures below 50\,K where the rates for low
initial $j$ values are enhanced by resonance contributions. This figure
also shows that our calculated total vibrational quenching rate agrees
with the most recent experimental data within the error bars, which
confirms the accuracy of our results.

The rovibrational transition rate coefficients presented in this paper
refer to vibrational de-excitation from $v = 1, \tilde{l}_v=1,j$ to
$v'=0,\tilde{l}'_v=0,j'$. The corresponding rovibrational excitation
rates can be easily obtained from the detailed balance relation
\cite{heijmen:99a}. In our previous papers on CO$_2$-He collisions with
CO$_2$ excited in the stretch modes \cite{selim:21,selim:22} we also
investigated the accuracy of cross sections and rate coefficients from
the multi-channel distorted-wave Born approximation (MC-DWBA) and the
nearest-neighbor Coriolis coupling (NNCC) method, which is an extension
of the CSA method, and we compared the computational efficiency of both
methods relative to CC. Our conclusion was that cross sections and rates
from the MC-DWBA method are practically equal to those from the full CC
method, or from the NNCC method when MC-DWBA was combined with the
latter method. The extension of CSA to NNCC resulted in better agreement
with full CC results. We did not investigate these methods in the
present paper on the bend mode of CO$_2$, but we expect similar results
in terms of accuracy and efficiency.

\section{Conclusions}
\label{sec:concl}

With the use of a newly computed four-dimensional CO$_2$-He potential
which includes the CO$_2$ bend coordinate we calculated the cross
sections $\sigma_{{v',l'_v,j' \leftarrow v,l_v,j}}(E)$ and rate
coefficients $k_{{v',l'_v,j' \leftarrow v,l_v,j}}(E)$ of rovibrational
transitions between different bend vibrational and rotational ($v,j$)
states of CO$_2$ induced by collisions with He atoms. The quantum number
$l_v$ represents the vibrational angular momentum generated by the bend
mode of CO$_2$, which is linear at equilibrium. In our scattering
calculations we used the numerically exact coupled-channels (CC) method,
but also the coupled-states approximation (CSA) and the vibrational
close-coupling rotational infinite-order sudden (VCC-IOS) approximation.
The effects of each of these approximations on the rovibrational cross
sections and rates was found to be less than 50\% at the rotational
state-to-state level, except for the smaller ones and in the low energy
resonance region, and less than 20\% on the overall vibrational
quenching rates, except for temperatures below 50\,K where resonances
provide a substantial contribution. Our calculated collisional
state-to-state transition rate coefficients can be used in modeling
interstellar non-LTE environments and our results show that they may
also be obtained from the computationally less demanding CSA or VCC-IOS
methods when the high accuracy achieved by CC calculations is not
required. Our CC quenching rates agree with the most recent experimental
data \cite{siddles:94} withing the error bars.

The fairly good performance of the VCC-IOS method seems surprising,
since we found in calculations on the symmetric and asymmetric stretch
modes of CO$_2$ \cite{selim:21} in collisions with He that the cross
sections and rate coefficients from VCC-IOS calculations differ by one
to three orders of magnitude from full CC results. The disagreement was
worse for the asymmetric stretch mode with frequency 2349 cm$^{-1}$ than
for the symmetric stretch mode at 1333 cm$^{-1}$ and the cross sections
were larger for the latter mode, so we concluded in Ref.~\cite{selim:21}
that the quality of VCC-IOS ameliorates with the magnitude of the cross
sections and we guessed already that VCC-IOS might perform better for
the bend mode at 667 cm$^{-1}$. Figure~S8 in the Supplement shows total
quenching cross sections from CC and VCC-IOS calculations of the bend
mode at collision energies 100 and 1000 cm$^{-1}$ calculated for
hypothetical frequencies in the range from 333 to 1500 cm$^{-1}$. This
figure confirms our expectations. It shows clearly that the VCC-IOS
cross sections agree fairly well with CC results up to a certain
frequency, but deviate more and more for higher frequencies. For
collision energy 100 cm$^{-1}$ this critical frequency is about 850
cm$^{-1}$, which is higher than the frequency of 667 cm$^{-1}$ of the
bend mode. For higher collision energies the cross section is larger and
the critical frequency, i.e., the energy gap between the $v=1$
and $v=0$ states where the VCC-IOS cross section starts deviating from
the CC result, increases; at 1000 cm$^{-1}$ it is about 1200 cm$^{-1}$.

We also compared our cross sections and rates for CO$_2$(bend)-He
collisions with data from Clary \textit{et al.}
\cite{clary:83,banks:87b} calculated in the 1980's with the CSA and
VCC-IOS methods and a simple atom-atom model potential based on
\textit{ab initio} Hartree-Fock calculations. Their cross sections agree
fairly well with ours for collision energies above 500 cm$^{-1}$ but at
lower energies the cross sections from our potential are much larger,
with the difference increasing to more than two orders of magnitude at
collision energies below 10~cm$^{-1}$. This shows that the inclusion of
long range attractive dispersion interactions is crucial to obtain
reliable cross sections at lower energies and rate coefficients at lower
temperatures.

Finally we note that the computer programs and the knowledge about the
accuracy of the approximate but computationally simpler CSA and VCC-IOS
methods will also be useful to provide similar collisional data of
astronomical interest for collisions of CO$_2$ with H$_2$ and for other
linear molecules of which bend mode spectra have been observed and
are used in modeling, such as C$_2$H$_2$ and HCN.

\section*{Data availability}
The data that support the findings of this study are available within
the article and the Supplementary material.

\section*{Supplementary material}
The Supplementary material contains Figs.~S1 to S8, which are discussed
in the main text. It also contains a copy of the Fortran program that
calculates the CO$_2$-He potential for CO$_2$ being deformed along
the bend mode normal coordinate.

\section*{Acknowledgements}
We thank Ewine van Dishoeck and Arthur Bosman for stimulating and useful
discussions. The work is supported by The Netherlands Organisation for
Scientific Research, NWO, through the Dutch Astrochemistry Network
DAN-II.

\clearpage


\end{document}


\title{State-to-state rovibrational transition rates for CO$_2$ in the bend mode in collisions with He atoms;
Supplementary material}

\author{Taha Selim} 
\email{tselim@science.ru.nl}
\affiliation{Theoretical Chemistry \\
Institute for Molecules and Materials, Radboud University \\
Heyendaalseweg 135, 6525 AJ Nijmegen, The Netherlands}

\author{Ad van der Avoird} 
\affiliation{Theoretical Chemistry \\
Institute for Molecules and Materials, Radboud University \\
Heyendaalseweg 135, 6525 AJ Nijmegen, The Netherlands}

\author{Gerrit C. Groenenboom}  
\email{gerritg@theochem.ru.nl}
\affiliation{Theoretical Chemistry \\
Institute for Molecules and Materials, Radboud University \\
Heyendaalseweg 135, 6525 AJ Nijmegen, The Netherlands}

\date{\today}

\maketitle

This Supplementary material contains Figs.~S1 to S8 mentioned in the paper.

\begin{figure}[ht!]
\includegraphics*[width=0.80\textwidth]{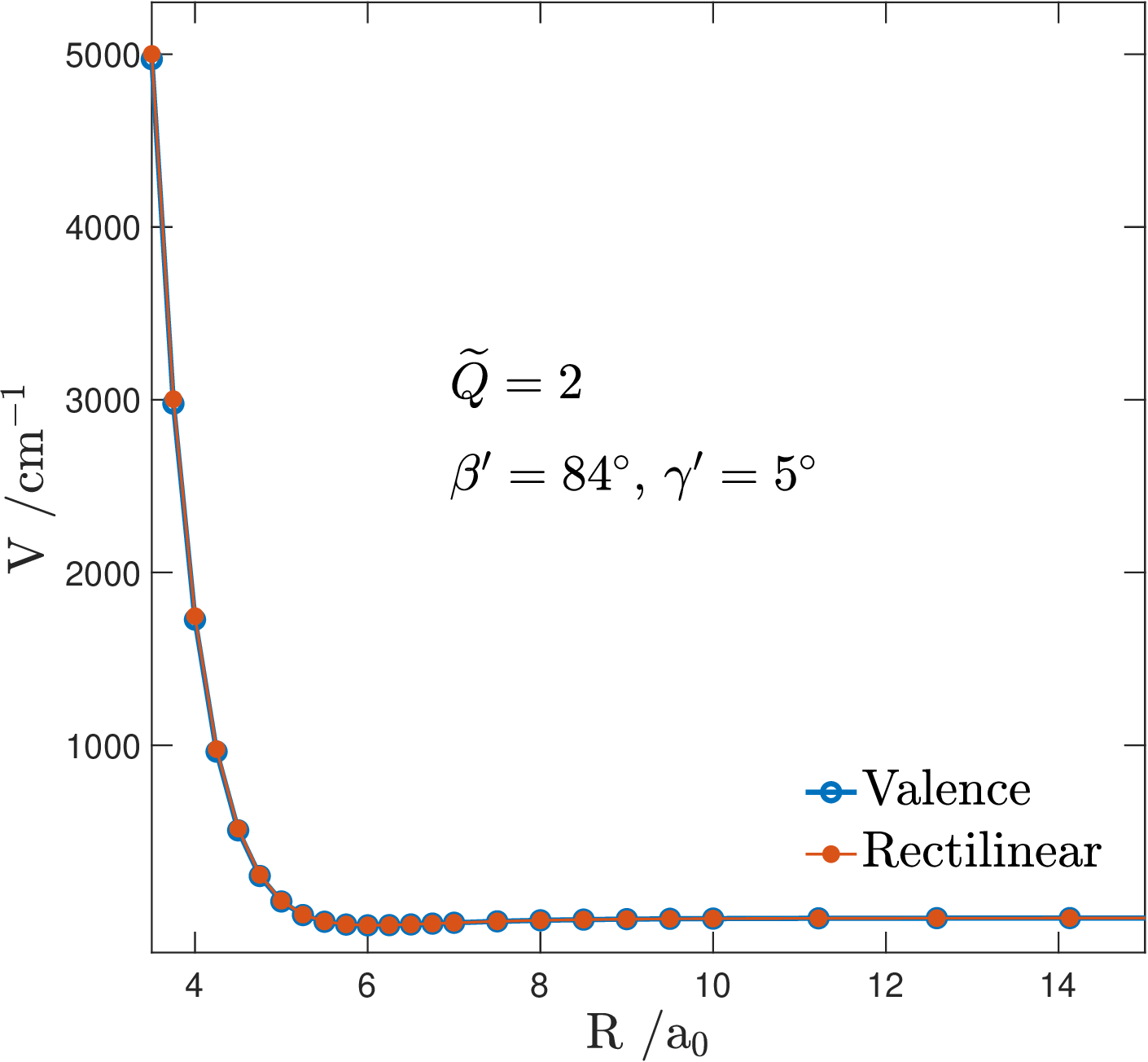}
\caption{
Radial cut of the 4D potential of CO$_2$-He at $\tilde{Q} = 2$ (O-C-O angle
$165^\circ$), $\beta' = 84^\circ$, and $\gamma' = 5^\circ$
calculated using a rectilinear or curvilinear normal coordinate $\tilde{Q}$.}
\label{fig:IPES_Qnorm_Qvalence}
\end{figure}

\begin{figure}[ht!]
\includegraphics*[width=0.80\textwidth]{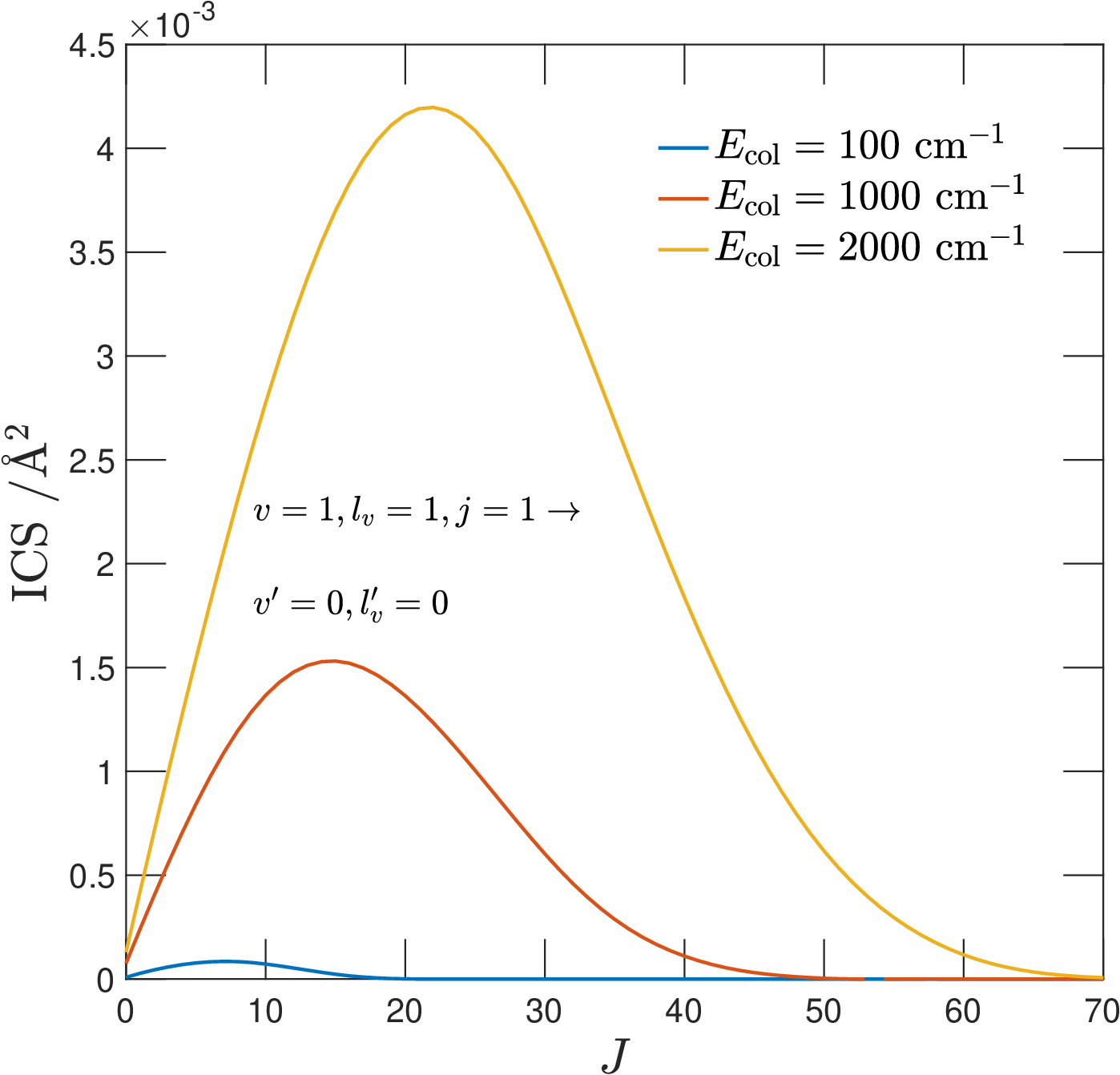}
\caption{
Partial wave distribution of the $v=1, \tilde{l}_v=1, j=1$ to
$v'=0, \tilde{l}'_v=0$ cross sections summed over all $j'$ and
both parities computed at the CC level.}
\label{fig:opacity}
\end{figure}

\begin{figure}[!ht]
\centering
\includegraphics[scale=0.48]{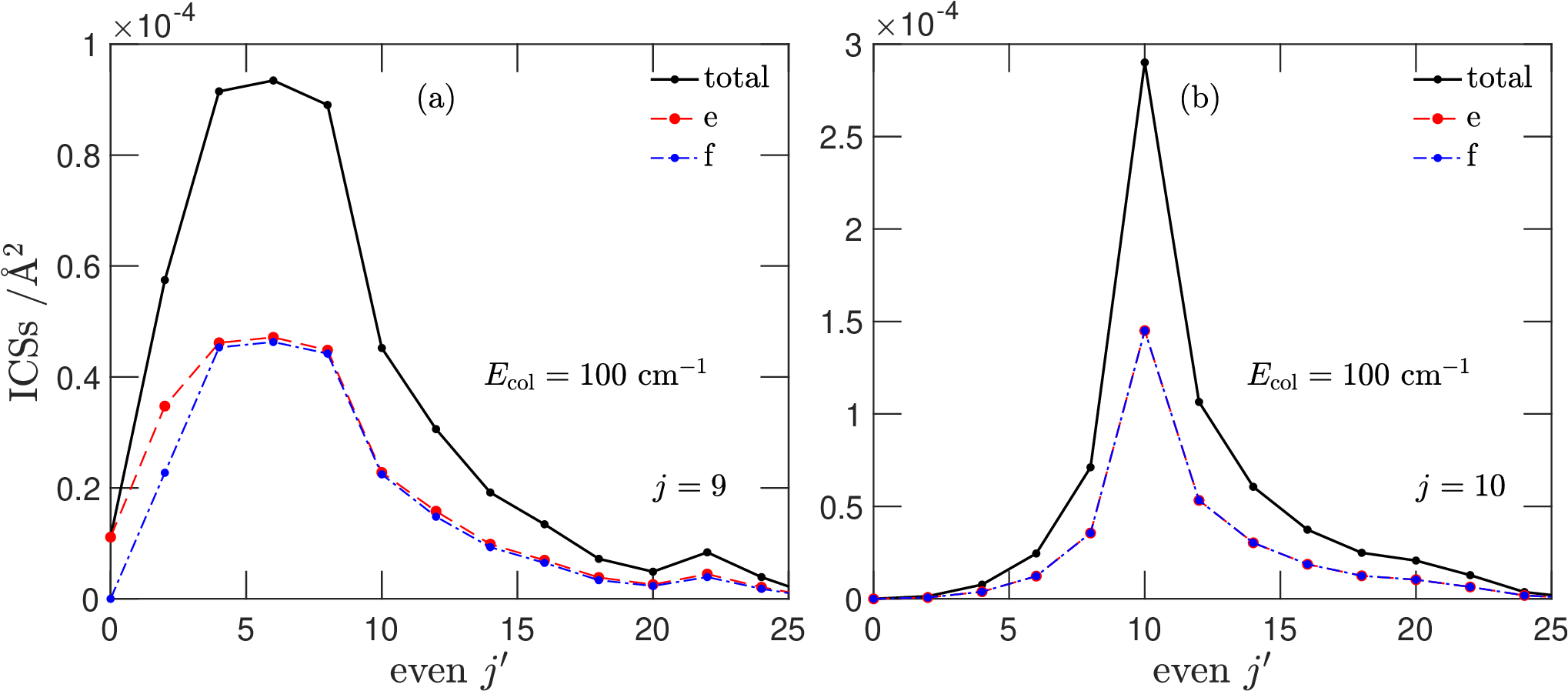}
\caption{Parity resolved state-to-state $v=1, \tilde{l}_v=1, j$ to
$v'=0, \tilde{l}'_v=0,j'$ cross sections for initial $j=9$ and 10 computed
at the CC level at collision energy $E=100$~cm$^{-1}$ resolved by
spectroscopic parity $e$ or $f$. The closed black curves are the total
cross sections, the dashed red and dot-dashed blue curves are
the contributions from parities $e$ and $f$, respectively. Note that for
initial $j=9$ the contributions from parity $e$ are larger, especially
for low $j'$, while for $j=10$ the contributions from parities $e$ and
$f$ are equal.}
\label{fig:sttost_parity}
\end{figure}

\begin{figure}[!ht]
\centering
\includegraphics[scale=0.5]{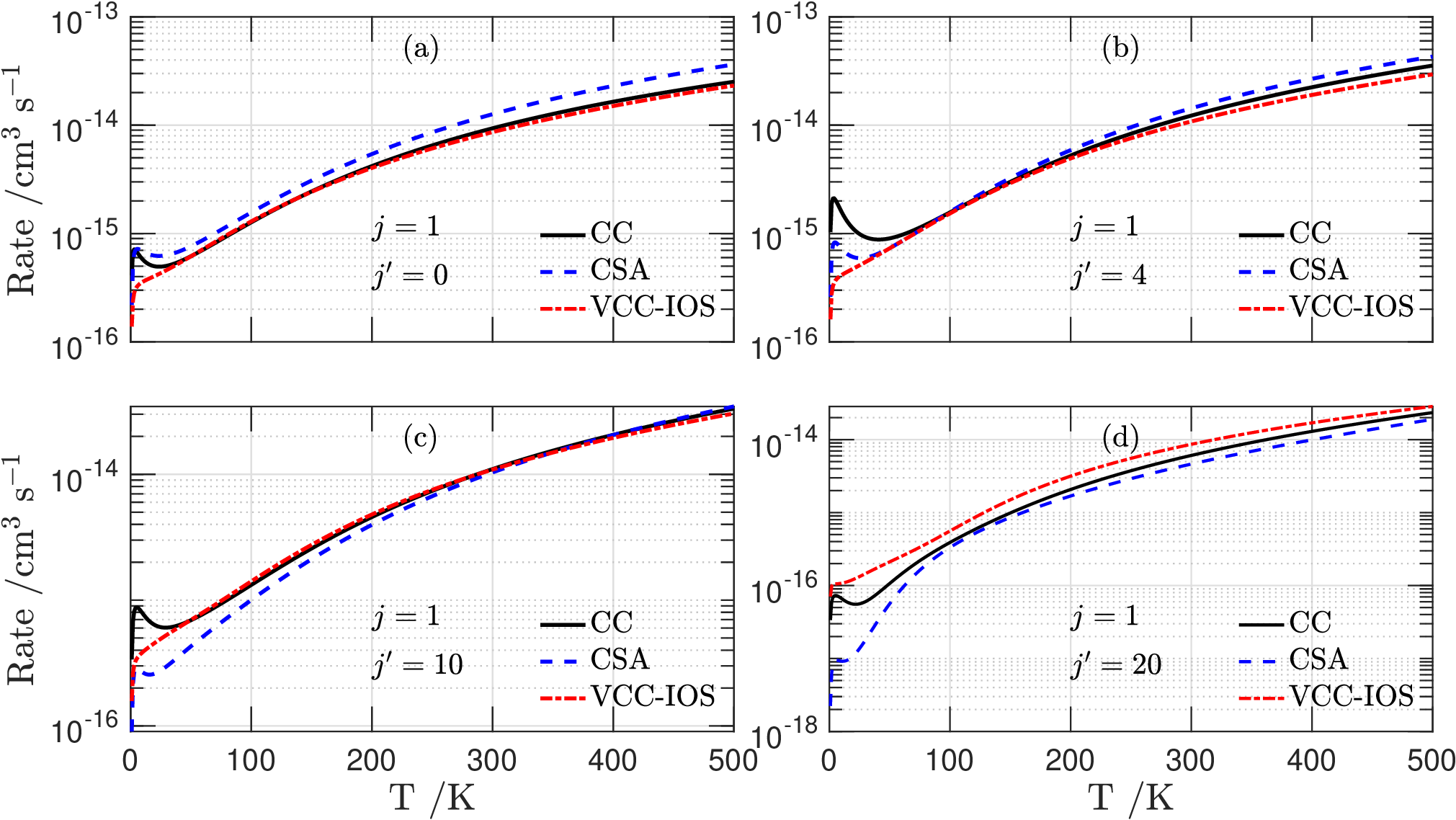}
\caption{State-to-states transition rate coefficients calculated with
different methods for initial $j=1$ and different final $j'$ values.}
\label{fig:rates_j1}
\end{figure}
\begin{figure}[!ht]
\centering
\includegraphics[scale=0.5]{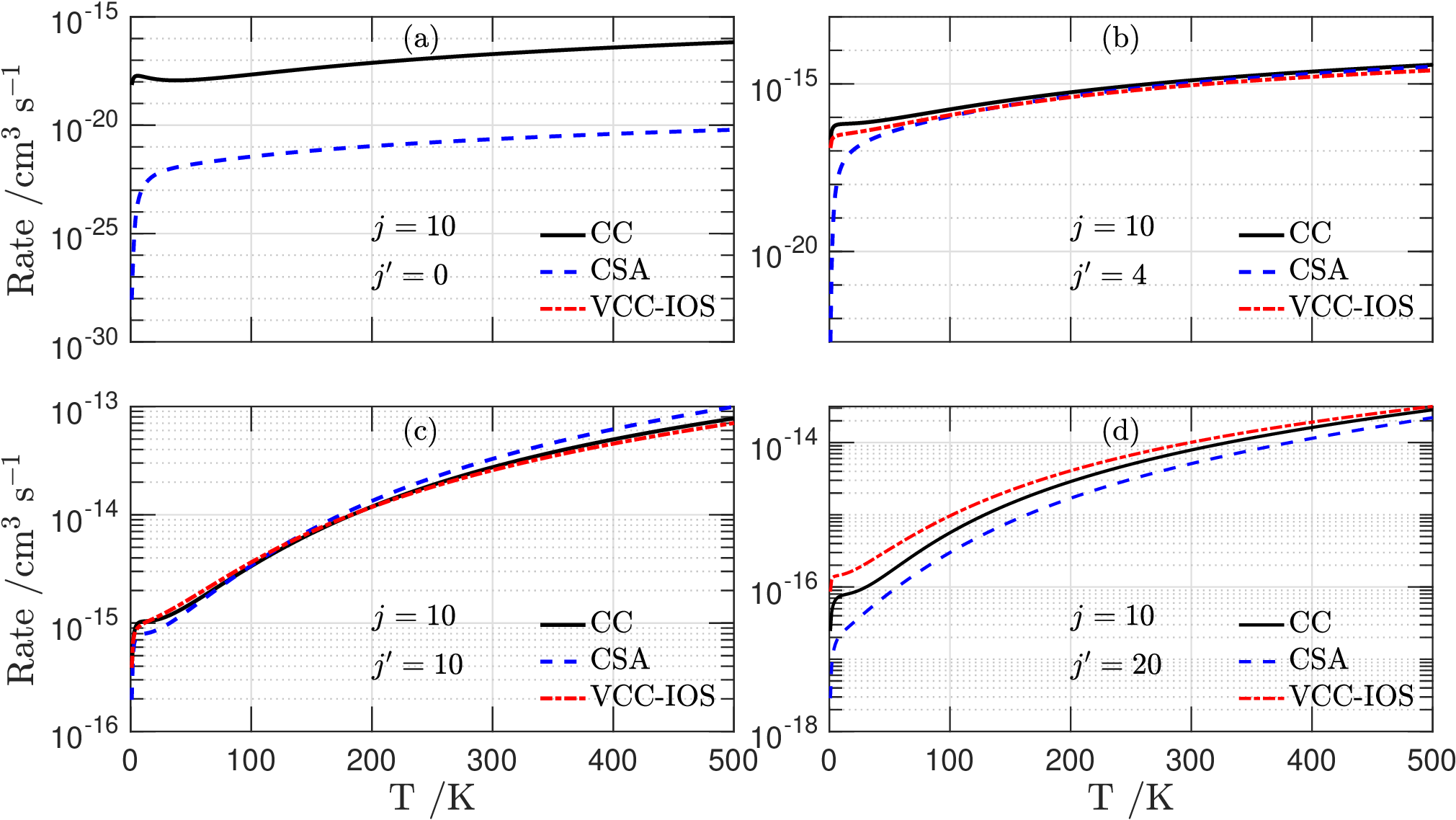}
\caption{Simlar to Fig.~\ref{fig:rates_j1} for initial $j=10$.}
\label{fig:rates_j10}
\end{figure}

\begin{figure}[!ht]
\centering
\includegraphics[scale=0.45]{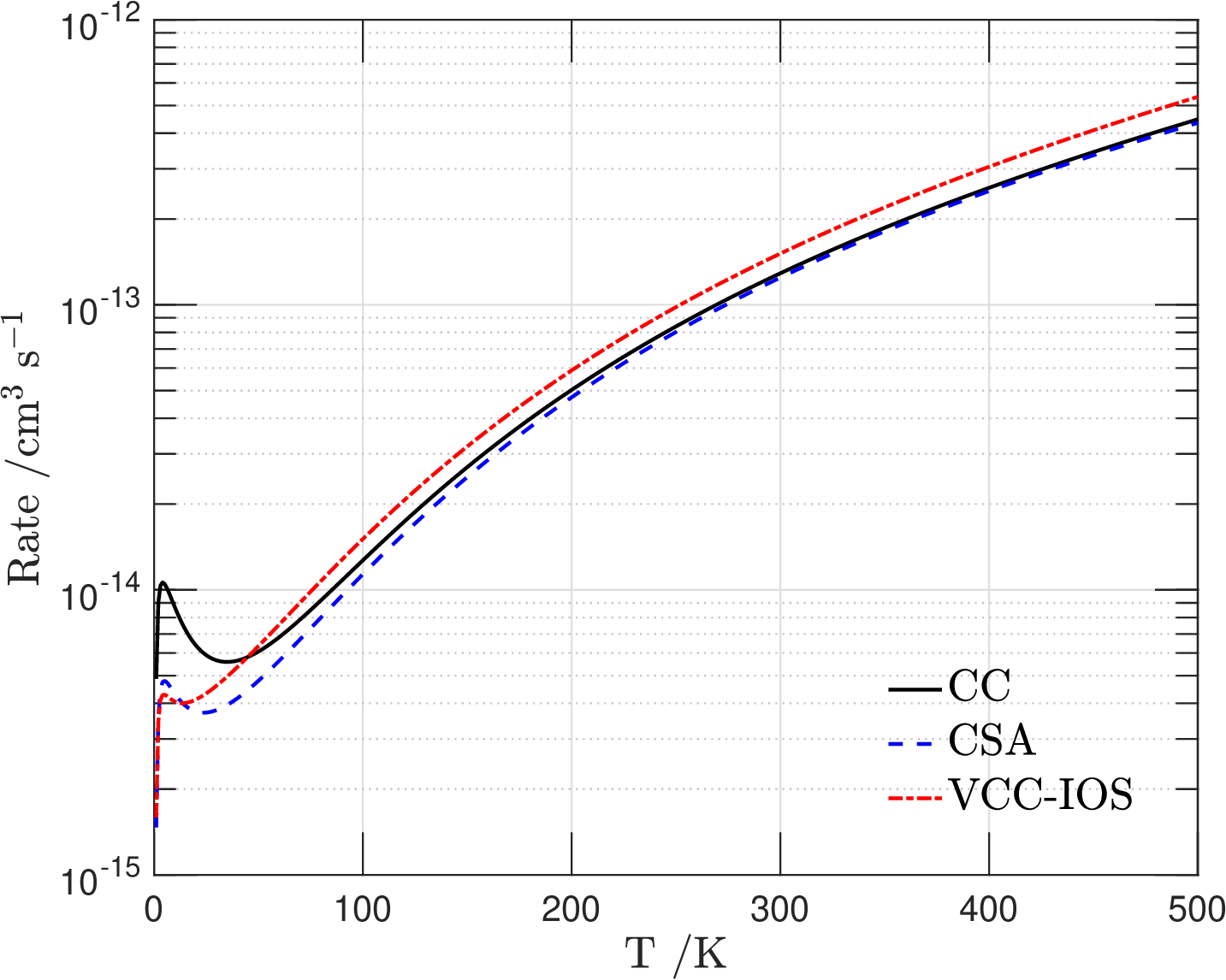}
\caption{Overall $v=1, \tilde{l}_v=1 \rightarrow v'=0, \tilde{l}'_v=0$
quenching rates transition rate coefficients calculated with
different methods for initial $j=1$.}
\label{fig:quench_j1}
\end{figure}
\begin{figure}[!ht]
\centering
\includegraphics[scale=0.45]{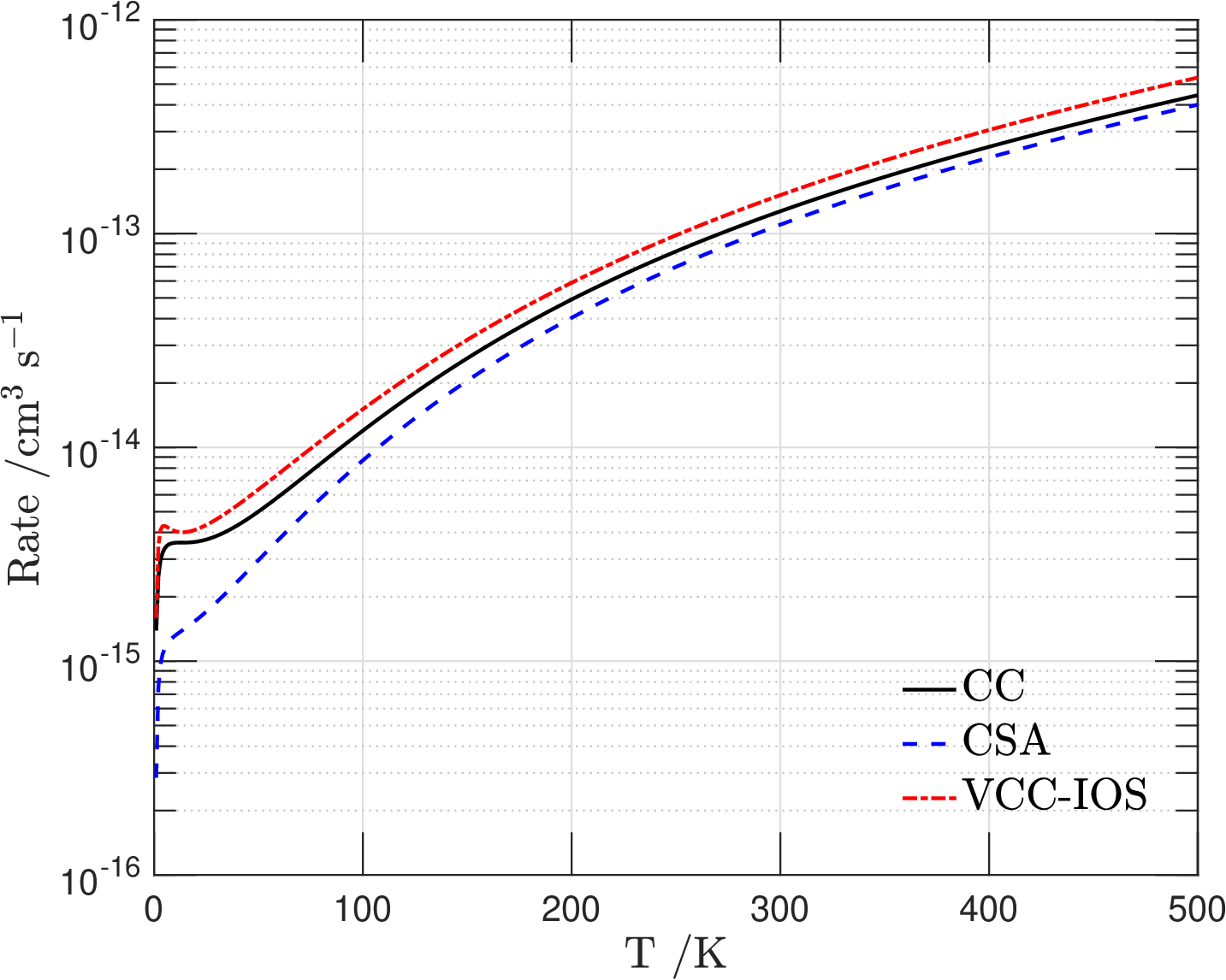}
\caption{Overall $v=1, \tilde{l}_v=1 \rightarrow v'=0, \tilde{l}'_v=0$
quenching rates transition rate coefficients calculated with
different methods for initial $j=10$.}
\label{fig:quench_j10}
\end{figure}

\begin{figure}[!ht]
\centering
\includegraphics[scale=0.5]{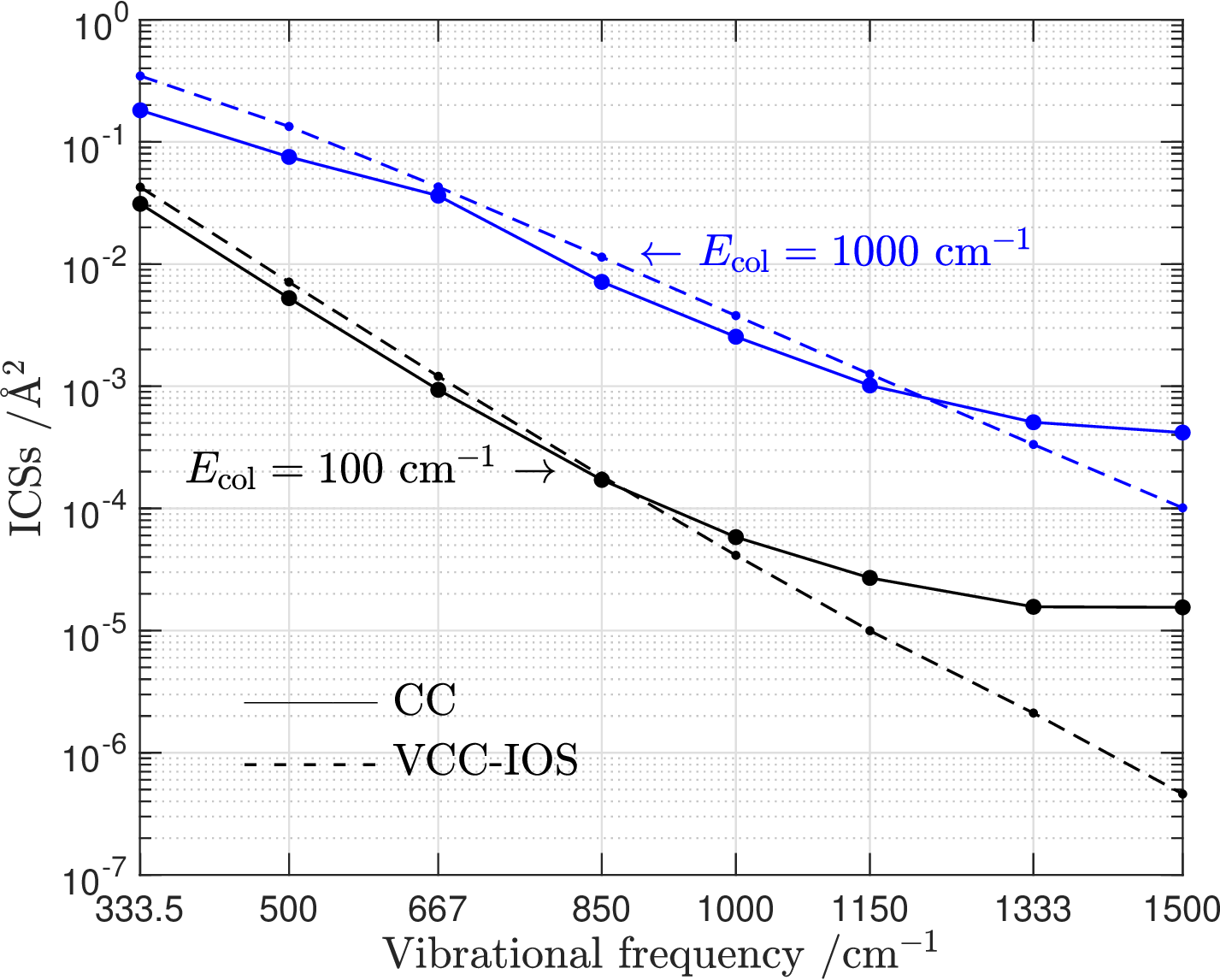}
\caption{Vibrational $v=1, \tilde{l}_v=1 \rightarrow v'=0, \tilde{l}'_v
= 0$ quenching cross sections sections from CC and VCC-IOS calculations
for the bend mode of CO$_2$ at collision energies 100 and 1000 cm$^{-1}$
calculated for hypothetical bend frequencies in the range from 333 to
1500 cm$^{-1}$. In the CC calculations the initial rotational state has
$j=1$ and the cross sections are summed over all final $j'$ states.}
\label{fig:vibfreq}
\end{figure}

\clearpage

\bibliography{../../vanderwaals}
\bibliographystyle{jcp}